\documentclass[10pt,a4paper]{article}

\pdfoutput=1
\usepackage{mathrsfs}

\usepackage{jheppub}

\usepackage{amsmath}
\usepackage{amssymb}
\usepackage{graphicx,color}

\usepackage{amsmath,amsthm}
\usepackage{amssymb}

\usepackage{graphicx}
\usepackage{bm}

\newcommand{\Ref}[1]{(\ref{#1})}

\newtheorem{Theorem}{Theorem}[section]
\newtheorem{Definition}{Definition}[section]

\newtheorem{Proposition}[Theorem]{Proposition}

\newcommand{\half}{\frac{1}{2}}

\newcommand{\Slc}{\mathrm{SL}(2,\mathbb{C})}

\newcommand{\Su}{\mathrm{SU}(2)}

\def\be{\begin{eqnarray}}
\def\ee{\end{eqnarray}}

\newcommand{\ca}{\mathcal A}

\newcommand{\cd}{\mathcal D}

\newcommand{\ck}{\mathcal K}
\newcommand{\cl}{\mathcal L}
\newcommand{\cm}{\mathcal M}

\newcommand{\cp}{\mathcal P}

\newcommand{\ct}{\mathcal T}

\newcommand{\cv}{\mathcal V}

\newcommand{\fg}{\mathfrak{g}}

\renewcommand{\a}{\alpha}
\renewcommand{\b}{\beta}
\newcommand{\g}{\gamma}
\newcommand{\G}{\Gamma}

\newcommand{\eps}{\varepsilon}

\newcommand{\sig}{\sigma}

\renewcommand{\l}{\lambda}
\renewcommand{\L }{\Lambda}

\renewcommand{\O}{\Omega}
\renewcommand{\t}{\tau}

\newcommand{\rmd}{\mathrm d}

\newcommand{\lt}{\left}
\newcommand{\rt}{\right}

\newcommand{\lag}{\left\langle}
\newcommand{\rag}{\right\rangle}

\newcommand{\tr}{\mathrm{tr}}

\newcommand{\fvol}{{}^4\cv}

\newcommand{\sD}{{\not}D}

\title{Spinfoam Fermions:\\
PCT Symmetry, Dirac Determinant, and Correlation Functions}

\author[a]{Muxin Han}

\author[a]{ \ Carlo Rovelli} 

\affiliation[a]{Centre de Physique Th\'eorique%
\footnote{Unit\'e mixte de recherche (UMR 6207) du CNRS et des Universit\'es de Provence (Aix-Marseille I), de la Meditarran\'ee (Aix-Marseille II) et du Sud (Toulon-Var); laboratoire affili\'e \`a la FRUMAM (FR 2291).}, CNRS-Luminy Case 907,  F-13288 Marseille, France}

\emailAdd{Muxin.Han-At-cpt.univ-mrs.fr}
\emailAdd{rovelli-At-cpt.univ-mrs.fr} 

\abstract{We discuss fermion coupling in the framework of spinfoam quantum gravity. We analyze the gravity-fermion spinfoam model and its fermion correlation functions. We show that there is a spinfoam analog of PCT symmetry for the fermion fields on spinfoam model, which is proved for spinfoam fermion correlation functions. We compute the determinant of the Dirac operator for the fermions, where two presentations of the Dirac determinant are given in terms of diagram expansions. We compute the fermion correlation functions and show that they can be given by Feynman diagrams on the spinfoams, where the Feynman propagators can be represented by a discretized path integral of a world-line action along the edges of the underlying 2-complex.  }


\begin{document}

\maketitle

\section{Introduction}

Loop Quantum Gravity (LQG) is an attempt to make a background independent, non-perturbative
quantization of 4-dimensional General Relativity (GR) -- for reviews, see \cite{book,rev,sfrevs}. It is inspired by the classical formulation of GR as a dynamical theory of connections. Starting from this formulation, the kinematics of LQG is well-studied and results in a successful kinematical
framework (see the corresponding chapters in the books \cite{book}). The framework of the dynamics in LQG is still largely open. There are two main approaches to the dynamics of LQG, they are (1) the Operator formalism of LQG, which follows the spirit of Dirac quantization or reduced phase space quantization of constrained dynamical system, and performs a canonical quantization of GR \cite{QSD}; (2) the Path integral formulation of LQG, which is currently understood in terms of the spin-foam formulation \cite{sfrevs,BC,EPRL,FK,HT}. The relation between these two approaches is well-understood in the case of 3-dimensional gravity \cite{perez}, while for 4-dimensional gravity, the situation is much more complicated and there are some recent attempts \cite{links} for relating these two approaches.

A serious shortcoming of LQG and the spin-foam models has long been the difficulty of coupling matter quantum field theory (see the first two references in \cite{sfrevs}), especially the coupling with fermions. It is still not clear so far about what is the behavior of the matter quantum fields on the quantum background described by LQG, and what are the quantum gravity corrections for matter quantum field theory. There was early pioneer works on coupling matter quantum field theory in canonical LQG \cite{matter} and in the context of spin-foam models in 3-dimensions and 4-dimensions e.g. \cite{3dfermion,FL,4dmatter}. And there was recent progress in \cite{SFfermion}, where we define a very simple form of fermion and Yang-Mills couplings in the framework of a 4-dimensional spin-foam formulation. Because of the simplicity of the fermion-coupling, it is possible for us to further analyze the detailed properties of the quantum fermion fields coupling to spin-foam quantum gravity.

In the present article, we mainly discuss the fermion coupling in the framework of 4-dimensional Lorentzian EPRL model \cite{EPRL}\footnote{The fermion coupling analyzed in this article can also be translated into Euclidean signature and implemented in the Euclidean EPRL-FK model and the model defined in \cite{HT}.}. The EPRL model in LQG is mostly inspired by the 4-dimensional Plebanski formulation of GR (Plebanski-Holst formulation by including the Barbero-Immirzi parameter $\b$), which is a BF theory constrained by the condition that the $B$ field should be ``simple'' i.e. there is a tetrad field $e^I$ such that $B={}^\star(e\wedge e)$. In the EPRL model, the implementation of simplicity constraint is understood in the sense of \cite{DingYou}. More importantly, the semiclassical limit of EPRL spin-foam model is shown to be well-behaved in the sense of \cite{semiclassical,holomorphic}.

Our analysis of fermion-coupling in this work follows the definition in \cite{SFfermion}. In Section \ref{GF}, we review the regularization procedure of the fermion action on a 2-complex $\ck$ and discuss its formal continuum limit. We also show that there is a way to express the Dirac fermion action (more precisely, the Dirac operator) in terms of spin-foam variables, so that the Dirac action is coupled into the spin-foam amplitudes. In this way we describe the dynamics of the fermion quantum field theory on a quantum background geometry, which is described by spin-foam model. Moreover we define and discuss the fermion correlation function on spin-foams. In \cite{3dfermion} it was mentioned that the (non-gauge-invariant) fermion correlation function vanishes on 3-dimensional spin-foam quantum gravity (the same thing also happens in lattice gauge theory, see the first reference of \cite{3dfermion} and the references therein), so one should make a certain gauge-fixing in order to define the correlation function properly. In 4-dimensions we would find the similar vanishing result for (non-gauge-invariant) fermion correlation function, if there was local $\Slc$ gauge invariance. However in Lorentzian spin-foam model for pure gravity, a gauge-fixing has been implemented in order to make the vertex amplitude finite \cite{finite}, such a gauge-fixing breaks the local $\Slc$ gauge invariance, and makes the fermion correlation functions well-defined. After that we discuss the PCT symmetry of the spin-foam fermions. The invariance under the inversions of charge, parity, and time simultaneously is believed to be a fundamental symmetry of nature. Here we define a transformation $\Theta$ for the gravity-fermion spin-foam partition function and its fermion correlation function, which can be viewed as a spin-foam analog of the PCT transformation in standard quantum field theory. Then a spin-foam PCT theorem is proved for the spin-foam fermion correlation functions, which states that the \emph{complex-conjugated} fermion correlation function on a spin-foam background equals the correlation function of \emph{charge-conjugated} fermions on a time-and-space reversed spin-foam background. This result is considered as a spin-foam analog of the celebrated PCT theorems proved for the quantum field theory on Minkowski spacetime \cite{PCT} and on curved spacetime \cite{hollands}.

In Section \ref{D}, we continue the computation for the gravity-fermion spin-foam model. If the integrations of the fermionic variables are carried out, it results in a determinant of the Dirac operator on the spin-foam model. This Dirac determinant contains the information about the interaction between the fermion field and gravitational field. So in Section \ref{D} we provide two representations for computing the spin-foam Dirac determinant in terms of diagrams.

In Section \ref{corr}, we compute the n-point correlation functions of spin-foam fermions. It turns out that the resulting spin-foam fermion correlation functions can be understood as coupling free-fermion Feynman-diagrams into the spin-foam amplitude, while each amplitude from a Feynman diagram depends on the spin-foam background geometry, which is summed over in the spin-foam amplitude. Because here we only consider the interaction between fermions and gravity, the Feynman diagrams coupled with the spin-foams are free fermion Feynman diagrams, which are completely factorized into Feynman propagators (matrix elements of inverse Dirac operator). And it turns out that the Feynman propagators can be expressed as a discretized path integration of a certain world-line action, where the world-lines are along the edges of the 2-complex underlying the spin-foam amplitudes. Our results confirm to some extend the early idea in \cite{FL} which proposes the inclusion of matter quantum fields by coupling their Feynman diagrams into spin-foam model.

\section{Gravity-Fermion Spin-foam Model and PCT Symmetry}\label{GF}

\subsection{Definition of Spin-foam Fermion}

Given a 2-complex $\ck$ dual to a simplicial complex $\Delta$, we consider the discretization on the complex $\ck$ the classical Dirac action:
\be
S_F:=\int_M \rmd^4x\ e\ \frac{i}{2}\lt[\overline{\psi}\g^\mu D_\mu\psi-\overline{D_\mu\psi}\g^\mu\psi\rt]-m_0\overline{\psi}\psi\label{dirac}
\ee
where $\g^\mu(x)=\g^Ie_I^\mu(x)$ is the spinorial tetrad, $e(x)=\det (e_\mu^I)$, and $D_\mu$ is the covariant derivative for Dirac spinor, i.e.
\be
D_\mu\psi=\partial_\mu\psi+\frac{1}{2}A_\mu^{IJ}S_{IJ}\psi
\ee
here $S^{IJ}:=\frac{1}{4}[\g^I,\g^J]$ is the Lie algebra generator of Lorentz group.

We first consider the first term in the action, it can be written as
\be
S_1:=\frac{i}{2}\int_M\overline{\psi}\g^I D\psi\wedge e^J\wedge e^K\wedge e^L\eps_{IJKL} \label{S1}
\ee
which motivate us to make the following anzatz for the naive discretization of $S_1$. We assign a fermion $\psi_v$ to each vertex $v\in V(\ck)$ of the complex $\ck$, and make the following formally discretized $S_1$ \cite{Ren,SFfermion}
\be
S_1\simeq 2i\sum_{e\in E(\ck)}V_e\ \overline{\psi_{b(e)}}\ \g^I\ {n_I(e)}\ \lt[G_{e}\psi_{f(e)}-\psi_{b(e)}\rt]\label{S11}
\ee
where $b(e)$ and $f(e)$ are respectively the begin and final point of $e$, $n^I(e)=\frac{\dot{e}^\mu(v)e_\mu^I(v)}{|\dot{e}^\mu(v)e_\mu^I(v)|}$ ($v=b(e)$) is a unit vector at the begin point of $e$, $V_e$ is a 3-volume associated to the edge $e$, which can be viewed as the volume of the tetrahedron (polyhedron) $\t_e$ dual to $e$ \footnote{Here we use $\t$ to denote both a tetrahedron and the center of a tetrahedron. We can make this notation because of their one-to-one correspondence.}, and
\be
G_e:=\cp e^{\frac{1}{2}\int_eA^{IJ}S_{IJ}}
\ee
is a $\Slc$ group element represented on the Dirac spinors.

When we check the formal continuum limit of Eq.\Ref{S11}, we consider a region $\O$, such that is much larger than the scale of an elementary cell (e.g. a 4-simplex if $\ck$ is dual to a simplicial complex), but smaller than the scale over which the fermion field and gravitational field (and their derivatives) change significantly. In this region $\O$, along each edge, ($\dot{e}_i^\mu=(\partial/\partial{s_i})^\mu$)
\be
G_{e_i}\psi_{f(e_i)}&\simeq&\Big[1+\Delta s_i\ \dot{e}_i^\mu\ \frac{1}{2}A^{IJ}_\mu S_{IJ}\Big]\Big[\psi_{b(e_i)}+\Delta s_i\ \dot{e}_i^\mu\ (\partial_\mu\psi)_{b(e_i)}\Big]\nonumber\\
&\simeq&\psi_{b(e_i)}+\Delta s_i\ \dot{e}_i^\mu\ (D_\mu\psi)_{b(e_i)}
\ee
Then at the region $\O$, the formal continuum limit of Eq.(\ref{S11}) given by
\be
2i\sum_{e\subset\O}V_e\ \overline{\psi}_{b(e)}\ {n}_I(e)\g^I\ \Delta s_e\ \dot{e}^\mu\ (D_\mu\psi)_{b(e)}
&=&2i\sum_{e\subset\O}V_e\ \overline{\psi}_{b(e)}\ {n}_I(e)\g^I\ \Delta s_e\ \dot{e}^\mu\ e^J_\mu e^\a_J(D_\a\psi)_{b(e)}\nonumber\\
&=&\frac{i}{2}\text{Vol}(\O) \overline{\psi}\g^\a D_\a\psi
\ee
where we have used the averaging formula:
\be
\sum_{e\subset\O}V_e{n}_I(e)\ \Delta s_e\ \dot{e}^\mu e^J_\mu=\frac{1}{4}\text{Vol}(\O)\delta_I^J
\ee
To illurstrate this averaging formula: firstly because there is a very large number of edges (with all the possible directions) in the region $\O$\footnote{One may also need to specify a distribution of links and their time-like/space-like nature in the case of Lorentzian signature, with some similar arguments as \cite{BHS}.}, the left hand side of this formula is invariant under 4-dimensional rotation, thus is proportional to $\delta^J_I$. Secondly, if we take the trace of the left hand side, it gives the volume of the region $\O$. Here we emphasize that the above argument of continuum limit is a formal (or naive) one, which help us choose a certain discretization of the fermion action. The true continuum limit of an interacting discrete quantum theory is a delicate issue. In the context of lattice QCD, the correct continuum limit has only been proved in the perturbation theory \cite{lattice}. The continuum limit of spinfoam formulation is currently an active research direction in LQG. The analysis of the continuum limit of spinfoam fermion coupling in quantum level is beyond the scope of the present paper.

Now we consider the discretized $S_1$ in Eq.\ref{S11}. The local $\Slc$ gauge transformations $U_v$ act as follows
\be
&&G_{e}\mapsto U_{b(e)}G_{e}U_{f(e)}^{-1}\ \ \ \ \ \psi_v\mapsto U_v\psi_v\ \ \ \ \ \overline{\psi}_v\mapsto\overline{\psi}_vU_v^{-1}\ \ \ \ \ n_I\mapsto\frac{1}{4}\tr\lt(\g_IU_v\g^JU_v^{-1}\rt)n_J\label{gauge}
\ee
The discretized $S_1$ is invariant under these gauge transformations. This can be seen by using the relation
\be
U_v\g^J U_v^{-1}=\g^I\frac{1}{4}\tr\lt(\g_IU_v\g^JU_v^{-1}\rt)
\ee
Similarly, the complex conjugate term $S_2$ in Eq.(\ref{dirac}) can be discretized similarly
\be
S_2&\simeq& -\frac{i}{2}\sum_{e\in E(\ck)}V_e\ \overline{\lt[G_e\psi_{f(e)}-\psi_{b(e)}\rt]}\ \g^I\ n_I(e)\ \psi_{b(e)}
\ee
while the mass term $S_3$ is given by
\be
S_3&\simeq&-m_0\sum_{v\in V(\ck)}\fvol_v\overline{\psi}_v\psi_v
\ee
where $\fvol_v$ is the 4-volume associate with the 4-simplex dual to $v$. The expression of 3-volume and 4-volume in terms of spinfoam variables are discussed in a short moment in the paragraph close to Eq.\Ref{4volume}.

As a result, the formally discretized action reads
\be
S_F[\psi_v,g_e]&\simeq&2i\sum_{e\in E(\ck)}V_e\lt[\overline{\psi}_{b(e)}\ \g^I\ n_I(e)\ G_{e}\psi_{f(e)}-\overline{\psi}_{f(e)}G^{-1}_e\ \g^I\ n_I(e)\ \psi_{b(e)}\rt]-m_0\sum_{v\in V(\ck)}\fvol_v\overline{\psi}_v\psi_v\nonumber\\
&\equiv&\sum_{e\in E(\ck)}S_e[\psi_{b(e)},\psi_{f(e)},g_e]+\sum_{v\in V(\ck)}S_v[\psi_v]
\ee
where $G_e$ is the representation of $g_e\in\Slc$ on Dirac spinors.

So far the unit vector $n^I(e)$ are located at the begin point $v$ of each edge $e$, and we assume $n^I(e)$ to be time-like and future-directed, which means that it can be transformed into $(1,0,0,0)$ by a proper orthochronous Lorentz transformation. We make a parallel transportation of $n^I(e)$ from the begin point of $e$ to a middle point $\t$, such that $n^I(e)$ is transformed into $n^I(\t)=\delta^I_0=(1,0,0,0)$, i.e. we consider a Lorentz transformation
\be
n^I(e)\frac{1}{4}\tr\lt(\g_IG_{v\t}\g^JG_{v\t}^{-1}\rt)=\delta_0^J\ \ \ \Rightarrow \ \ \ \ n_I(e)\frac{1}{4}\tr\lt(\g^IG_{v\t}\g_JG_{v\t}^{-1}\rt)=n_I(e)\frac{1}{4}\tr\lt(\g_JG_{v\t}^{-1}\g^IG_{v\t}\rt)=\delta^0_J
\ee
while
\be
\L^I_{\ J}=\frac{1}{4}\tr\lt(\g_JG_{v\t}^{-1}\g^IG_{v\t}\rt)\ \ \ \Rightarrow \ \ \ \
(\L^{-1})^I_{\ J}=\frac{1}{4}\tr\lt(\g_JG_{v\t}\g^IG_{v\t}^{-1}\rt)
\ee
As a result
\be
n_I(e)=\frac{1}{4}\tr\lt(\g_IG_{v\t}\g^JG_{v\t}^{-1}\rt)\delta_J^0
\ee
then
\be
\g^In_I(e)=G_{v\t}\g^JG_{v\t}^{-1}\delta^0_J=G_{v\t}\g^0G_{v\t}^{-1}\label{paralleltrans}
\ee
We define the Lorentz transformation $G_{v\t}$ as the spin-foam $\Slc$ holonomy $g_{v\t}$ represented on the space of Dirac spinors. Recall that $n^I=(1,0,0,0)$ is the unit vector orthogonal to all the face bivecors of the tetrahedron (polyhedron) $\t$ by the simplicity constraint \cite{EPRL,DingYou,semiclassical}. So this definition states that the internal vector $n^I(e)$ coming from the tangent vector of the edge $e$ is the normal of the tetrahedron (polyhedron) $\t_e$ viewed in the frame at the vertex $v$. Therefore under this definition the fermion action is expressed by
\be
S_e&=&2i V_e\lt[\overline{\psi}_{b(e)}\ G_{b(e)\t_e}\g^0 G_{\t_ef(e)}\psi_{f(e)}-\overline{\psi}_{f(e)}G_{f(e)\t_e}\g^0G_{\t_eb(e)}\ \psi_{b(e)}\rt]\nonumber\\
S_v&=&-m_0\fvol_v\overline{\psi}_v\psi_v
\ee
We can also write $S_e$ in terms of Weyl spinors. In Weyl basis
\be
\psi=\left(\begin{array}{c}
                                 \xi^A   \\
                                 \theta_{A'} \\
                               \end{array}
                             \right)\ \ \ \ \ \ \ \g^0=\left(\begin{array}{cc}
                                 0\ & 1  \\
                                 1\ & 0 \\
                               \end{array}
                             \right)
\ee
A representation of $\Slc$ on Dirac spinors can thus be written as a tensor product between an $\Slc$ representation on 2-spinors $\xi^A$ and a $\Slc$ representation on dual complex-conjugated 2-spinors $\theta_{A'}$ \footnote{Given $g\in\Slc$ a $2\times2$ complex matrix with a unit determinant, its representation on 2-spinors is given by $\xi^A\mapsto g^A_{\ B}\xi^B$, its representation on dual complex-conjugated 2-spinors is given by $\theta_{A'}\mapsto -\overline{g}{}_{A'}^{\ B'}\theta_{B'}=-\eps_{C'A'}\overline{g}{}^{C'}_{\ D'}\eps^{B'D'}\theta_{B'}$ (from $\theta^{A'}\mapsto \overline{g}{}^{A'}_{\ B'}\theta^{B'}$).}. Then the action can be written as
\be
S_e
&=&2i V_e\lt[{\xi}_{b(e)}^\dagger g_{\t_e b(e)}^\dagger g_{\t_ef(e)}\xi_{f(e)}+{\theta}_{b(e)}^\dagger \overline{g}{}_{\t_e b(e)}^\dagger \overline{g}_{\t_ef(e)}\theta_{f(e)}-{\xi}_{f(e)}^\dagger g_{\t_e f(e)}^\dagger g_{\t_eb(e)}\xi_{b(e)}-{\theta}_{f(e)}^\dagger \overline{g}{}_{\t_e f(e)}^\dagger \overline{g}_{\t_eb(e)}\theta_{b(e)}\rt]\nonumber\\
S_v&=&-m_0\fvol_v\lt[\xi^\dagger_v\theta_v+\theta^\dagger_v\xi_v\rt]
\ee
where $g_{ve}$ is a $\Slc$ group element represented by $2\times 2$ complex matrix with unit determinant. Here $^\dagger$ denote complex conjugate and transpose\footnote{Using spinor language, $^\dagger$ for $\xi^A$ is defined by $\xi^\dagger_A=\overline{\xi}{}^{A'}\delta_{A'A}$, for $\theta_{A'}$ is defined by $(\theta^\dagger)^{A'}=\delta^{A'A}\overline{\theta}_A$.}. In terms of spinor indices
\be
{\xi}_1^\dagger g_1^\dagger g_2\xi_2&\equiv&\delta_{B'B}(\overline{g}_1)^{B'}_{\ A'}(g_2)^{B}_{\ A}(\overline{\xi}{}_1)^{A'}(\xi_2)^A\nonumber\\
{\theta}_1^\dagger \overline{g}{}_{1}^\dagger \overline{g}_{2}\theta_{2}&\equiv&\delta^{BB'}(g_1)_{B}^{\ A}(\overline{g}_2)_{B'}^{\ A'}(\overline{\theta}_1)_A(\theta_2)_{A'}\ =\ \delta_{BB'}(g_1)^{B}_{\ A}(\overline{g}_2)^{B'}_{\ A'}(\overline{\theta}_1)^A(\theta_2)^{A'}.
\ee
Note that similar to Eq.(\ref{paralleltrans}), we also have
\be
\sig_{A'A}^In_I(e)=\delta_I^0\sig^I_{B'B}(\overline{g}_{\t_e,b(e)})^{B'}_{\ A'}(g_{\t_e,b(e)})^{B}_{\ A}\ \ \ \ \text{and}\ \ \ \ \ \sig^0_{B'B}=\frac{1}{\sqrt{2}}\delta_{B'B}.
\ee
In the discretized fermion action, the 3-volume $V_e$ is a function of the spins $j_f$ and the coherent intertwiner $i_e$ \cite{poly}. More explicitly, we can express the vertex amplitude $A_v[j_f,i_e,g_{v e}]_{GR}$ in terms of Livine-Speziale coherent intertwiners $i_e=||\vec{j}_f,\vec{n}_{ef}\rangle$ \cite{LS}, whose labels $(\vec{j}_f,\vec{n}_{ef})$ determines the geometry of a tetrahedron (or polyhedron) \cite{poly} where the spins $j_f$ and unit vectors $\vec{n}_{ef}$ satisfies the closure condition $\sum_f j_f \vec{n}_{ef}=0$ \footnote{Then the sum over all the intertwiners in the spin-foam model will be a integral over all the unit 3-vectors $n_{ef}$ satisfying the closure constraint \cite{poly}.}. So we write the 3-volume $V_e$ as a function of the labels $(\vec{j}_f,\vec{n}_{ef})$ in the case of a tetrahedron \cite{poly}. 
\be
V_e=\frac{\sqrt{2}}{3}\sqrt{\lt|\vec{A}_{f_1}\cdot\lt(\vec{A}_{f_2}\times \vec{A}_{f_3}\rt)\rt|},\ \ \ \ \vec{A}_{f}\equiv \g j_f\vec{n}_{ef}.
\ee
For the 4-volume $\fvol_v$ of a 4-simplex, we can write it as a function of the boundary data $(\vec{j}_f,\vec{n}_{ef})$ and $\Slc$ group element $g_{ve}$ by the following procedure. Given a unit 4-vector $(1,0,0,0)$, we specify a so(3) sub-algebra inside the Lorentz Lie algebra (as a real Lie algebra), while we denote by the $4\times4$ matrices $L_i$ $(i=1,2,3)$ the so(3) generators in the Lorentz Lie algebra (in its vector representation). Given a tetrahedron dual to an edge $e$, its face bivectors are given by $B_{e f}=j_f n_{ef}^iL_{i}$. This bivector viewed from the frame of 4-simplex is defined by a parallel transportation $B_{vf}=\fg_{ve}B_{ef}\fg_{ev}=j_f n_{ef}^i\fg_{ve}L_{i}\fg_{ev}$ where $\fg$ is the vector representation of $g\in\Slc$. Given two triangles $f,f'$ which don't belong to the same tetrahedron, we can write a 4-volume
\be
\fvol_v(f,f')=\prec B_{vf},B_{vf'}\succ=j_f j_{f'} n_{ef}^i n_{e'f'}^k \prec\fg_{ve}L_{i}\fg_{ev},\fg_{ve'}L_{k}\fg_{e'v}\succ\label{4volume}
\ee
where the inner product $\prec B_1,B_2\succ:=\eps_{IJKL}B_1^{IJ}B_2^{KL}$. The above way to write 4-volume depends on the choice of two triangles $f,f'$, thus the expression of $\fvol_v$ should be the above quantity $\fvol_v(f,f')$ averaged by all the possible choices of $f,f'$. However in the large-j regime of EPRL model \cite{semiclassical} or the model in \cite{HT}, where the shape-matching condition is implemented, $\fvol_v(f,f')$ doesn't depend on the choice of $f,f'$, because the simplicity condition $B_{ef}\cdot (1,0,0,0)^t=0$ has been implemented strongly, and the simplicity, closure, and shape-matching conditions implies the face bivectors $B_{vf}$ can be expressed in terms of the wedge produces of cotriads \cite{BJ,HZ}.

We have expressed all the quantities in the fermion action in terms of spin-foam variables. Let's now couple the fermions to spin-foam quantum gravity. The spin-foam model for pure gravity on a 2-complex $\ck$ with a boundary graph $\g$ \cite{EPRL,FK,KKL}
\be
Z_{GR}(\ck)_f=\sum_{j_f}\int \rmd\mu_{j_f}(i_e)\int_{\Slc}\rmd g_{v e}\prod_{f} A_f[j_f]_{GR}\prod_{v}A_v[j_f,i_e,g_{v e}]_{GR}\cdot f_{\g,j_f,i_e}
\ee
where $f_{\g,j_f,i_e}$ is a boundary state on the boundary graph $\g$ in its spin-network representation. $\rmd\mu_{j_f}(i_e)=\rmd\mu_{j_f}(\vec{n}_{ef})$ is an integral measure for the coherent intertwiner $i_e=||\vec{j}_f,\vec{n}_{ef}\rangle$ with fixed $j_f$, defined in the first reference of \cite{poly}, such that the integral $\int \rmd\mu_{j_f}(i_e)$ only integrates over the unit vectors $\vec{n}_{ef}$ satisfy the closure condition $\sum_f j_f \vec{n}_{ef}=0$. Given a tetrahedron $t_e$, the measure $\rmd\mu_{j_f}(i_e)$ can be written as 
\be
\rmd\mu_{j_f}(\vec{n}_f)=\prod_{f\subset t_e}\rmd^2\vec{n}_f\delta^{(3)}\lt(\sum_f j_f \vec{n}_{f}\rt)\det\lt(G(\vec{n}_{f})\rt)\int_{H^3}\prod_f\lt|\rho_h(\vec{n}_{f})\rt|^{2(j_f+1)}\rmd h
\ee
where the factor $\lt|\rho_h(\vec{n}_{f})\rt|^{2}$ is given by (the spin-1/2 coherent state $|n_f\rangle$ is a normalized 2-component spinor)
\be
\lt|\rho_h(\vec{n}_{f})\rt|^{2}=\lag n_f|h^\dagger h|n_f\rag^{-1}
\ee
The intgral over $H^3=\Slc/\Su$ describes an averaging of the coherent intertwiner along the action of $H^3$. The matrix $G(\vec{n}_{f})$ is the metric on the obits of the action
\be
G(\vec{n}_{f})_{ij}=\sum_{f\subset t_e}j_f\lt(\delta^{ij}-n_f^in_f^j\rt).
\ee

In order to couple the Dirac fermion to gravity, we add an edge amplitude to $Z_{GR}(\ck)_f$
\be\label{Z}
&&Z(\ck):=\sum_{j_f}\int \rmd\mu_{j_f}(i_e)\int_{\Slc}\rmd g_{v e}\int[\cd\psi_v\cd\overline{\psi}_v]\prod_{f} A_f[j_f]\prod_{v}A_v[j_g,i_e,g_{ve},\psi_v]\prod_eA_e[\psi_{b(e)},\psi_{f(e)},g_{ve},j_f,i_e]\cdot f_{\g,j_f,i_e}\nonumber\\
&&\text{where}\ \  A_f[j_f]=A_f[j_f]_{GR},\ \ \ A_e[\psi_{b(e)},\psi_{f(e)},g_{ve},j_f,i_e]=e^{iS_e},\ \ \ A_v[j_g,i_e,g_{ve},\psi_v]=A_v[j_g,i_e,g_{ve}]_{GR}\ e^{iS_v}
\ee
The measure $[\cd\psi_v\cd\overline{\psi}_v]$ is defined by the Grassmann integral
\be
[\cd\psi_v\cd\overline{\psi}_v]=\prod_{(v,\a)}\rmd{\psi}^\a_v\rmd\overline{\psi}{}^\a_v.
\ee

Note that here we follow the spinfoam formulation of the structure defined by the EPRL and FK models \cite{EPRL,FK}, with the generalization by \cite{KKL,DingYou} to arbitrary 2-complex\footnote{The massive fermion coupling to spinfoam is defined on a 2-complex dual to simplicial complex, since the mass term involve the 4-volume, which is the 4-volume of 4-simplex when we define it using spinfoam variable. If the fermion is massless, the coupling can be made on an arbriary 2-complex.}. Importantly the factor $\prod_{v}A_v[j_f,i_e,g_{v e}]_{GR}$ in these models (either in Euclidean or Lorenzian signature) can be written as either directly $e^S$ or $\int e^S\rmd\mu$ with some additional integration variables \cite{semiclassical,CF,HZ} if the intertwiners $i_e$ are the Livine-Speziale coherent intertwiners. Therefore the pure gravity spinfoam amplitude can be written as a path integral form respecting a discrete ``spinfoam action'' $S$. This spinfoam action is critical in the recent semiclassical analysis of spinfoam formulation \cite{semiclassical,CF,HZ,BJ}. Therefore here the coupling of fermion with spinfoam quantum gravity can be understood as a coupling in the level of action, i.e. the gravity-fermion spinfoam amplitude can be expressed in a path integral form with respect to an action $S+S_e+S_v$ where $S$ is the gravity part and $S_e+S_v$ is the fermion part.

\subsection{Fermion Correlation Functions}

With the gravity-fermion spin-foam model defined above, we consider a correlation function with a number of $\overline{\psi}_v$ and $\psi_v$ insertions:
\be
&&\lag\overline{\psi}_{v_1}\cdots\overline{\psi}_{v_n}\psi_{v_{n+1}}\cdots\psi_{v_{n+m}}\rag_{\text{Spinfoam}}\nonumber\\
&:=&\sum_{j_f}\int \rmd\mu_{j_f}(i_e)\int_{\Slc}\rmd g_{ve}\prod_{f}A_f[j_f]_{GR}\prod_{v}A_v[j_f,i_e,g_{ve}]_{GR}\cdot f_{\g,j_f,i_e}\times\nonumber\\
&&\ \ \ \ \ \times\int[\cd\psi_v\cd\overline{\psi}_v]\ \overline{\psi}_{v_1}\cdots\overline{\psi}_{v_n}\psi_{v_{n+1}}\cdots\psi_{v_{n+m}}\ \exp\Big(iS_F\lt[\ck,j_f,i_e,g_{ve},\psi_v\rt]\Big)\label{correlator}
\ee
The above definition of correlation function needs some explanations: Given that (1) the vertex amplitude $A_v[j_f,i_e,g_{ve}]_{GR}$ is invariant under the $\Slc$ gauge transformation $g_{ve}\mapsto \l_vg_{ve}$ ($\l_v\in\Slc$) at the vertex $v$; (2) the fermion action $S_F\lt[\ck,j_f,i_e,g_{ve},\psi_v\rt]$ is invariant under the $\Slc$ gauge transformation $G_{ve}\mapsto \L_vG_{ve}$, $\psi_v\mapsto\L_v\psi_v$ and $\overline{\psi}_v\mapsto\overline{\psi}_v\L^{-1}_v$ ($\L_v$ and $G_{ve}$ are the $\Slc$ elements $\l_v$ and $g_{ve}$ represented on the space of Dirac spinors); (3) the measures $\rmd g_{ve}$ and $\rmd\psi_v\rmd\overline{\psi}_v$ is invariant under the $\Slc$ gauge transformation, we make a change of variables $g_{ve}\mapsto g_{ve}^\l= \l_vg_{ve}$, $\psi_v\mapsto\psi_v^\l=\L_v\psi_v$ and $\overline{\psi}_v\mapsto\overline{\psi}{}_v^\l=\overline{\psi}_v\L^{-1}_v$ of the integral (which doesn't change the integral at all), then use the gauge invariance of the vertex amplitude, fermion action, and the measures
\be
&&\lag\overline{\psi}_{v_1}\cdots\overline{\psi}_{v_n}\psi_{v_{n+1}}\cdots\psi_{v_{n+m}}\rag_{\text{Spinfoam}}\nonumber\\
&=&\sum_{j_f}\int \rmd\mu_{j_f}(i_e)\int_{\Slc}\rmd g^\l_{ve}\prod_{f}A_f[j_f]_{GR}\prod_{v}A_v[j_f,i_e,g^\l_{ve}]_{GR}\cdot f_{\g,j_f,i_e}\times\nonumber\\
&&\ \ \ \ \ \times\int[\cd\psi^\l_v\cd\overline{\psi}{}^\l_v]\ \overline{\psi}{}^\l_{v_1}\cdots\overline{\psi}{}^\l_{v_n}\psi^\l_{v_{n+1}}\cdots\psi^\l_{v_{n+m}}\ \exp\Big(iS_F\lt[\ck,j_f,i_e,g^\l_{ve},\psi^\l_v\rt]\Big)\nonumber\\
&=&\sum_{j_f}\int \rmd\mu_{j_f}(i_e)\int_{\Slc}\rmd g_{ve}\prod_{f}A_f[j_f]_{GR}\prod_{v}A_v[j_f,i_e,g_{ve}]_{GR}\cdot f_{\g,j_f,i_e}\times\nonumber\\
&&\ \ \ \ \ \times\int[\cd\psi_v\cd\overline{\psi}_v]\ \overline{\psi}{}^\l_{v_1}\cdots\overline{\psi}{}^\l_{v_n}\psi^\l_{v_{n+1}}\cdots\psi^\l_{v_{n+m}}\ \exp\Big(iS_F\lt[\ck,j_f,i_e,g_{ve},\psi_v\rt]\Big)
\ee
Therefore naively one would have
\be
&&\lag\overline{\psi}_{v_1}\cdots\overline{\psi}_{v_n}\psi_{v_{n+1}}\cdots\psi_{v_{n+m}}^{}\rag_{\text{Spinfoam}}
\ =\ \lag\overline{\psi}{}^\l_{v_1}\cdots\overline{\psi}{}^\l_{v_n}\psi^\l_{v_{n+1}}\cdots\psi^\l_{v_{n+m}}\rag_{\text{Spinfoam}}\nonumber\\
&&=\ \L_{v_{n+1}}\cdots\L_{v_{n+m}}\lag\overline{\psi}_{v_1}\cdots\overline{\psi}_{v_n}\psi_{v_{n+1}}\cdots\psi_{v_{n+m}}^{}\rag_{\text{Spinfoam}}
\L^{-1}_{v_1}\cdots\L^{-1}_{v_n}
\ee
which results in that if $v_1,\cdots,v_{n+m}$ are different vertices and the correlation function is non-gauge-invariant, then it vanishes because the gauge transformations $\l_v$ are independent at different vertices. The same result was mentioned in \cite{3dfermion} in the context of 3-dimensional gravity, where it was suggested that one should either make a gauge-fixing to define the non-gauge-invariant correlation function, or instead consider the correlation functions of gauge invariant quantities, e.g. the fermion currents.

However the above argument didn't take into account the regularization of the Lorentzian vertex amplitude \cite{finite}. If we consider pure gravity amplitude only, and our vertex amplitude $A_v[j_f,i_e,g^\l_{ve}]_{GR}$ has the gauge invariant under $g_{ve}\mapsto \l_vg_{ve}$, then the $\Slc$ integrals
\be
\int_{\Slc}\prod_{e\ \text{at}\ v}\rmd g_{ve}\ A_v[j_f,i_e,g_{ve}]_{GR}
\ee
is divergent, because we can always choose an edge $e_0$ at $v$ and a gauge $\l_v=g_{ve_0}^{-1}$, such that $g_{ve}\mapsto g_{ve_0}^{-1}g_{ve}$ ($e\neq e_0$) and $g_{ve_0}\mapsto 1$, then there is a redundant $\Slc$ integral $\int\rmd g_{ve_0}$ which gives the divergence. The same argument for divergence also applies to the gravity-fermion spin-foam model by the gauge invariance of the fermion action $S_F\lt[\ck,j_f,i_e,g_{ve},\psi_v\rt]$. The way to remove the divergence is firstly to choose an edge $e_0$ at each vertex $v$, and fix the gauge $\l_v=g_{ve_0}^{-1}$ at each vertex, such that $g_{ve}\mapsto g_{ve_0}^{-1}g_{ve}$ ($e\neq e_0$) and $g_{ve_0}\mapsto 1$, so the integrand doesn't depend on the variable $g_{ve_0}$ for each vertex. Then the redundant integral $\int\rmd g_{ve_0}$ should be removed at each vertex. It turns out that at least for the pure gravity spin-foam amplitude, the $\Slc$ integrals lead to finite result after regularization for a large class of spin-foam vertices \cite{finite}.

Therefore the definition of fermion correlation function Eq.(\ref{correlator}) should be understood in terms of the gauge-fixed/regularized spin-foam amplitude, where there is a edge $e_0$ at each vertex $v$ such that the $\Slc$ group element $g_{ve_0}$ is gauge fixed to be the identity $1\in\Slc$, and the integral of $g_{ve_0}$ is removed at each vertex $v$. The previous argument for vanishing correlation function doesn't apply for the gauge-fixed/regularized spin-foam amplitude because the gauge-fixing breaks the $\Slc$ gauge invariance. Thus the correlation function Eq.(\ref{correlator}) is \emph{not} necessary vanishing. More explicitly we could write the correlation function in the following way instead of Eq.\Ref{correlator}
\be
&&\lag\overline{\psi}_{v_1}\cdots\overline{\psi}_{v_n}\psi_{v_{n+1}}\cdots\psi_{v_{n+m}}\rag_{\text{Spinfoam}}\nonumber\\
&:=&\sum_{j_f}\int \rmd\mu_{j_f}(i_e)\int_{\Slc}\rmd g_{ve}\prod_{f}A_f[j_f]_{GR}\prod_{v}A_v[j_f,i_e,g_{ve}]_{GR}\cdot f_{\g,j_f,i_e}\times\prod_{(v,e_0)}\delta_{\Slc}(g_{ve_0})\times\nonumber\\
&&\ \ \ \ \ \times\int[\cd\psi_v\cd\overline{\psi}_v]\ \overline{\psi}_{v_1}\cdots\overline{\psi}_{v_n}\psi_{v_{n+1}}\cdots\psi_{v_{n+m}}\ \exp\Big(iS_F\lt[\ck,j_f,i_e,g_{ve},\psi_v\rt]\Big)\label{correlator1}
\ee
More detailed computation on the fermion correlation function will be given in Section \ref{corr}.

\subsection{PCT Symmetry on Spin-foam}

We first briefly recall the notion of PCT symmetry for Dirac fermion on flat and curved spacetime. Given the fermion field operator $\psi(t,\vec{x})$ on Minkowski spacetime, its Parity-Inversion, Time-Reversal, and Charge-Conjugation are defined by
\be
P\psi(t,\vec{x})P^{-1}=\g^0\psi(t,-\vec{x})\ \ \ \ \ T\psi(t,\vec{x})T^{-1}=-\g^1\g^3\psi(-t,\vec{x})\ \ \ \ \ C\psi(t,\vec{x})C^{-1}=-i\g^2\psi^*(t,\vec{x})
\ee
Thus the anti-unitary PCT transformation acts on the fermion field operator by (here we use $\psi^*$ to denote complex conjugation and $\psi^C$ to denote charge-conjugation)
\be
\psi^{C}(t,\vec{x})=(PCT)\psi(t,\vec{x})(PCT)^{-1}=(-i)\g^5\psi^*(-t,-\vec{x})=(-i)\g^5\g^0[\overline{\psi}(-t,-\vec{x})]^t
\ee
Then the PCT theorem on Minkowski spacetime states that: Given the Minkowski vacuum state $\O$, a class of local fields $\Phi(x)$ (can be composite fields) with $m$ primed spinor indices, and an anti-unitary PCT operator $(PCT)$
\be
(PCT)\Phi(x)(PCT)^{-1}=\Phi^C(x)=(-1)^m(-i)^F\Phi(-x)^*,\ \ \ \ \ \ (PCT)\O=\O
\ee
where $F=0$ if $\Phi$ is bosonic field and $F=1$ if $\Phi$ is fermionic field, the PCT theorem on Minkowski spacetime states that the complex conjugated correlation function of a number of fields equals the correlation function of the corresponding charge-conjugated fields \cite{PCT}, i.e.
\be
\lag\O\big|\Phi_1^{C}(x_1)\cdots\Phi_n^{C}(x_n)\big|\O\rag=\lag\O\big|\Phi_1(x_1)\cdots\Phi_n(x_n)\big|\O\rag^*.
\ee

On Minkowski spacetime the map $\rho: (t,\vec{x})\mapsto(-t,-\vec{x})$ defines an isometry preserving the spacetime orientation but reversing the time orientation. For a general (globally hyperbolic) spacetime $\cm$ with metric $g_{\a\b}$ and time orientation and spacetime orientation $o=(T,\eps_{\a\b\g\delta})$, we let $\overline{\cm}$ be the spacetime with the identical manifold structure and metric structure as $\cm$, but its time orientation and spacetime orientation is given by $-o=(-T,\eps_{\a\b\g\delta})$. Because there is no preferred vacuum state on a general curved spacetime, the PCT theorem on a general spacetime \cite{hollands} is formulated in terms of Operator-Product-Expansion (OPE) coefficients: Given a local field $\Phi(x)$ with $n$ unprimed spinor indices and $m$ primed spinor indices, we define its charge-conjugated field by
\be
\Phi^C(x):=(-1)^m(-i)^F\Phi(x)^*
\ee
Here $x$ denotes the points on the spacetime manifold, and $\Phi^C(x)$ is a local field with $m$ unprimed spinor indices and $n$ primed spinor indices. We suppose the a class of fields $\Phi^{(j)}_\cm(x)$ ($j=1,\cdots,n$) on the spacetime $\cm$ have the following OPE in short geodesic distance
\be
\Phi^{(1)}_\cm(y_1)\cdots\Phi^{(n)}_\cm(y_n)\sim\sum_{(j)}c_{\cm,x}^{(j)}(y_1,\cdots,y_n)\ \Phi^{(j)}_\cm (x)
\ee
as $(y_1,\cdots,y_n)$ approaching $x$, where the distributions $c_{\cm,x}^{(j)}(y_1,\cdots,y_n)$ are OPE structure coefficients. Then the PCT theorem on a general curved spacetime implies that on the spacetime $\overline{\cm}$ with the opposite time and space orientation, the OPE of the charge-conjugated fields $\Phi^{(j)C}_{\overline{\cm}(x)}$ ($j=1,\cdots,n$) is given by \cite{hollands}
\be
\Phi^{(1)C}_{\overline{\cm}}(y_1)\cdots\Phi^{(n)C}_{\overline{\cm}}(y_n)\sim\sum_{(j)}c_{\cm,x}^{(j)}(y_1,\cdots,y_n)^*\ \Phi^{(j)C}_{\overline{\cm}} (x)
\ee
whose OPE structure coefficients are the complex conjugation of $c_{\cm,x}^{(j)}(y_1,\cdots,y_n)$.

In the formalism of spin-foam model, the quantization of spacetime $\cm$ with metric $g_{\a\b}$ is formulated by a spin-foam amplitude $Z(\ck)$ on a 2-complex $\ck$. The reversal of time orientation while keeping spacetime orientation unchange $o\mapsto -o$ can be formulated by simultaneously reversing all the internal edge orientations in the complex $\ck$. The reason is the following: Given a spacetime $(\cm,g_{\a\b})$ with space and time orientation $o=(T,\eps_{\a\b\g\delta})$. All the oriented orthonormal frames are given by $e^\a_I$, satisfying
\be
g_{\a\b}e^\a_Ie^\b_J=\eta_{IJ}\ \ \ \ \ e_0^\a\nabla_\a T>0\ \ \ \ \ \eps_{\a\b\g\delta}\ e_0^\a\ e^\b_1\ e^\g_2\ e^\delta_3>0
\ee
The oriented orthonormal frames form the frame bundle $F(\cm)$ over $\cm$ whose structure group is the proper orthochronous Lorentz group. However for the spacetime $(\overline{\cm},g_{\a\b})$ with opposite space and time orientation $-o=(-T,\eps_{\a\b\g\delta})$, its frame bundle $F(\overline{\cm})$ is naturally isomorphic to $F(\cm)$ by the map
\be
I:\ F(\cm)\to F(\overline{\cm}),\ \ \ \ e_I^\a\mapsto -e^\a_I
\ee
We consider the spin-foam model defined on a 2-complex $\ck$ dual to a simplicial complex imbedded in the spacetime manifold. Given a tetrahedron $\t$ in the simplicial complex, all the area bivectors of the tetrahedron are orthogonal to a unit internal vector $n^I=(1,0,0,0)$ by the simplicity constraint \cite{EPRL,DingYou,semiclassical}. And this vector $n^I$ is given by the tangent vector of the edge $e$ dual to the tetrahedron $\t$ up to a proper orthochronous Lorentz transformation, i.e. on the spacetime manifold $\cm$ with orientation $o$
\be
a\L^I_{\ J}n^J=\dot{e}^\a e_\a^I\ \ \ \ \ (a>0)
\ee
where $\L^I_{\ J}$ is a proper orthochronous Lorentz transformation. However if the spin-foam model is build on the spacetime manifold $\overline{\cm}$ with orientation $-o$ (in another words, if the spin-foam model is a quantization of the spacetime structure $(\cm,g_{\a\b},-o)$), the previous co-frame field $e_\a^I$ changes into $-e_\a^I$, thus one has to change the previous tangent vector $\dot{e}^\a$ into $-\dot{e}^\a$, i.e. reverse the edge orientation, to keep $n^I$ unchange. Note that $n^I$ has to be fixed to be $(1,0,0,0)$ for the definition of spin-foam model.

We first formulate the PCT invariance of the spin-foam fermions in the following formal way: We define an anti-unitary PCT operator $\theta$, such that it acts on fermion field operators by
\be
\theta\psi_v\theta^{-1}=(-i)\g^5\g^0\overline{\psi}_v^t\ \ \ \ \ \theta\overline{\psi}_v\theta^{-1}=(-i)\psi_v^t\g^0\g^5
\ee
which are charge-conjugate field operators. If we consider the fermion action $S_F(\ck)$ (defined on a complex $\ck$) is an composite operator from the fermion field operator $\psi_v$ and $\overline{\psi}_v$, $S_F$ is PCT invariant in the sense that
\be
\theta S_F(\ck)\theta^{-1}=S_F(\ck^{-1})
\ee
It is indeed the case. We consider the bilinear form $2iV_e\overline{\psi}_v G_{ve}\g^0 G_{ev'}\psi_{v'}$ appearing in the discretized fermion action, by using $\{\g^5,\g^0\}=0$, $(\g^0)^2=(\g^5)^2=1$ and $(\g^0)^\dagger=\g^0=(\g^0)^t$, as well as the assumption $\theta$ is anti-unitary
\be
&&\theta\lt[2iV_e\overline{\psi}_v G_{ve}\g^0 G_{ev'}\psi_{v'}\rt]\theta^{-1}\nonumber\\
&=& -2iV_e (-i)\psi_v^t\g^0\g^5 {G}^*_{ve}\g^0 G^*_{ev'}(-i)\g^5\g^0\overline{\psi}_{v'}^t\
=\ -2iV_e\psi_v^t\g^0 {G}^*_{ve}\g^0 {G}^*_{ev'}\g^0\overline{\psi}_{v'}^t\
=\ -2iV_e\psi_v^t{G}^t_{e v}\g^0 {G}^t_{v'e}\overline{\psi}_{v'}^t\nonumber\\
&=& 2iV_e\overline{\psi}_{v'} G_{v'e}\g^0 G_{ev}\psi_{v}
\ee
where we have treated $V_e$ and $G_{ve}$ as c-numbers (gravity as external field), and we also used the fact that $\g^5$ commutes with $G$ and the relation $\g^0{G}^*\g^0=({G}^{* -1})^\dagger= (G^{-1})^t$. Here we have shown that the action of $\theta$ on the bilinear form interchanges the vertices $v$ and $v'$. Hence under this transformations of variables, $S_e[\psi_{b(e)},\psi_{f(e)}]$ transforms to
\be
S_e[\psi_{b(e)},\psi_{f(e)}]&\mapsto&2i V_e\lt[\overline{\psi}_{f(e)}\ G_{f(e)\t_e}\g^0 G_{\t_eb(e)}\psi_{b(e)}-\overline{\psi}_{b(e)}G_{b(e)\t_e}\g^0G_{\t_ef(e)}\ \psi_{f(e)}\rt]\nonumber\\
&&=\ S_{e^{-1}}[\psi_{b(e^{-1})},\psi_{f(e^{-1})}]
\ee
where we use that $b(e)=f(e^{-1})$ and $f(e)=b(e^{-1})$. On the other hand it is easy to see that the mass term $S_v=-m_0\fvol_v\overline{\psi}_v\psi_v$ is invariant under $\theta$. Hence we obtain the PCT invariance of the spin-foam fermion in the sense of
\be
\theta S_F(\ck)\theta^{-1}=S_F(\ck^{-1}).
\ee

Now we consider the gravity-fermion spin-foam model. We define the following transformation $\Theta$ of the spin-foam amplitude, which is an analog of PCT transformation:

\begin{Definition}
Given a 2-complex $\ck$ with a boundary graph $\g$, and the gravity-fermion spin-foam
\be
Z(\ck)_f&=&\sum_{j_f}\int \rmd\mu_{j_f}(i_e)\int_{\Slc}\!\!\!\rmd g_{ve}\prod_{f}A_f[j_f]_{GR}\prod_{v}A_v[j_f,i_e,g_{ve}]_{GR}\cdot f_{\g,j_f,i_e}\times\nonumber\\
&&\ \ \ \ \ \times\int[\cd\psi_v\cd\overline{\psi}_v]\ \exp\Big(iS_F\lt[\ck,j_f,i_e,g_{ve},\psi_v\rt]\Big)
\ee
where $f_{\g,j_f,i_e}$ is a boundary state in its spin-network representation, we define a spin-foam analog of PCT transformation $\Theta$ by
\begin{itemize}
  \item $\Theta$ reverses the orientations of all the internal edges in the complex $\ck$, i.e. $\Theta:\ck\mapsto\ck^{-1}$;
  \item $\Theta$ changes all the gravity vertex amplitudes $A_v[j_f,i_e,g_{ve}]_{GR}$ into their complex conjugates $A_v[j_f,i_e,g_{ve}]_{GR}^*$;
  \item The boundary state $f_{\g,j_f,i_e}$ transforms into its complex conjugate $f_{\g,j_f,i_e}^*$;
  \item $\Theta$ changes the fermion action on the exponential $iS_F(\ck)$ into
  \be
  \theta \lt[iS_F(\ck)\rt]\theta^{-1}=-iS_F(\ck^{-1});\label{-i}
  \ee
  \item For each fermion $\psi_v$ at a vertex, $\Theta:\psi_v\mapsto\psi^C_v:=(-i)\g^5\psi_v^*=(-i)\g^5\g^0\overline{\psi}_v^t$ (charge-conjugation), or in terms of Weyl spinors:
      \be
      \Theta:\ \left(\begin{array}{c}
                                 \xi_v   \\
                                 \theta_v \\
                               \end{array}
                             \right)\mapsto(-i)\left(\begin{array}{c}
                                 {\xi}^*_v   \\
                                 -{\theta}^*_v \\
                               \end{array}
                             \right).
      \ee
\end{itemize}
\end{Definition}

The second and third transformations are motivated by the anti-unitarity of the PCT operator. We consider heuristically the pure gravity spin-foam amplitude as a physical inner product between some certain in-state and out-state $f_{in}$ and $f_{out}$, which describe the boundary data of quantum gravity
\be
Z_{GR}(\ck)_f=\lag f_{out}\ ,\ f_{in} \rag_{Phys}
\ee
Then heuristically the PCT transformation reverses the in-state and out-state by
\be
\Theta: \lag f_{out}\ ,\ f_{in} \rag_{Phys}\mapsto \lag \theta f_{out}\ ,\ \theta f_{in} \rag_{Phys}=\lag f_{out}\ ,\ f_{in} \rag_{Phys}^*=\lag f_{in}\ ,\  f_{out}\rag_{Phys}
\ee
where we implicitly use the fact that the path integral measure $\rmd g_{ve}$ is real, since the $\Slc$ Haar measure $\rmd g$ can be written explicitly by
\be
\rmd g=\frac{\rmd\b\rmd{\b}^*\rmd\g\rmd{\g}^*\rmd\delta\rmd{\delta}^*}{|\delta|^2}\ \ \ \ \ g=\left(
                                                                                       \begin{array}{cc}
                                                                                        {\a} & \b \\
                                                                                        \g & \delta \\
                                                                                       \end{array}
                                                                                     \right)
\ee
which is manifestly invariant under $g\mapsto g^*$. The other integration measure $\rmd\mu_{j_f}(i_e)$ is a real measure \cite{poly}. 

Here we argue that the PCT transformation of pure gravity spin-foam amplitude
\be
\Theta Z_{GR}(\ck)_f={Z_{GR}(\ck^{-1})}_f^*
\ee
is a spin-foam analog of the spacetime $\overline{\cm}$ with reversed time and space orientation $-o=(-T,\eps_{\a\b\g\delta})$. We consider the semiclassical behavior of the pure gravity spin-foam amplitude. For a given 4-simplex, the corresponding vertex amplitude $A_v[j_f,i_e]_{GR}$ can be represented in terms of Livine-Speziale (LS) coherent intertwiners \cite{LS}, i.e. use LS coherent intertwiner $||{j}_f,{n}_f\rangle$ for the SU(2) intertwiner $i_e$. Thus we can denote the vertex amplitude by
\be
A_v[j_f,n_{ef}]_{GR}=\int\prod_{e\ at\ v}\rmd g_{ve} A_v[j_f,n_{ef},g_{ve}]_{GR}
\ee
where $n_{ef}\in S^2$ is a unit 3-vector associated to each triangle/face $f$. The large-j asymptotics of the vertex amplitude $A_v[j_f,n_{ef}]_{GR}$ was studies in \cite{semiclassical} when $j_f,n_{ef}$ satisy the closure condition: If $j_f$ goes to be large
\be
A_v[j_f,n_{ef}]_{GR}\sim N_+ e^{i\sum_f\b j_f\vartheta_f}+N_-e^{-i\sum_f\b j_f\vartheta_f}\label{asymp}
\ee
where $N_\pm$ are independent of $j_f$, and $\vartheta_f$ is the extrinsic angle between the two tetrahedra sharing the triangle $f$, i.e. for two tetrahedra $\t,\t'$ sharing a triangle $f$, $n^I(\t)$ and $n^I(\t')$ are two unit 4-vectors respectively orthogonal to all the face bivectors of $\t$ and $\t'$. The extrinsic angle $\vartheta_f$ is defined by $\cosh\vartheta_f=-\eta_{IJ}n^I(\t)n^J(\t')$. Then
\be
\sum_f\b j_f\vartheta_f[j_f]=S_{Regge}[j_f]
\ee
is the Regge action in a single 4-simplex with triangle areas $\ca_f=8\pi G\hbar\b j_f$ ($\b$ is the Barbero-Immirzi parameter).

The vertex amplitude can also be written in the holomorphic representation \cite{holomorphic} with boundary states to be the complexifier coherent states. Given the boundary graph $\g$ of the vertex amplitude, and for a given link $l$, we associate a complexifier coherent state $\psi^t_{g_l}(h_l)$ \cite{GCS} where
\be
\psi^t_{g}(h)=\sum_{j}(2j+1)e^{-j(j+1)\frac{t}{2}}\chi^j(gh^{-1})
\ee
where $t$ is a dimensionless classicality proportional to $\ell_p^2$, and $g_l\in\Slc$ is the complexified phase space coordinate $g_l=h_l e^{{E_l}t/{\ell_p^2\b}}$. These coherent states (with all possible $g_l$) form a over-complete basis of the Hilbert space $L^2(\text{SU(2)})$. The representation of the vertex amplitude on the complexifier coherent states is called the holomorphic representation of the vertex amplitude, denoted by $A_v[g_f,g_{ve}]_{GR}$ and is regarded as an amplitude with boundary data $g_f$ (a link $l$ of the boundary graph uniquely corresponds to a face $f$). In the holomorphic representation and as the limit $\ell_p\to0$ \cite{holomorphic,GCS}
\be
A_v[g_f]_{GR}\sim \sum_{j_f}A_v[j_f,n_{ef}]_{GR}\ \exp\lt[-i\sum_f j_f\xi_f\rt]\ \exp\lt[-\sum_f\lt(j_f-\frac{\ca_f}{\ell^2_p\b}\rt)^2\frac{t}{2}+\sum_f\lt(\frac{\ca_f}{\ell_p^2\b}\rt)^2\frac{t}{2}\rt]
\ee
where $\ca_f$ is the area of the triangle $f$ evaluated at the phase space point $g_f$, and $\xi_f=\b\vartheta$ is the extrinsic angle evaluated at the phase space point $g_f$. From this we see that as the limit $\ell_p\to0$, the large spins $j_f\sim\frac{\ca_f}{\ell^2_p\b}$ dominate the contributions of $A_v[g_f,g_{ve}]_{GR}$. We then consider the large spin contributions and insert in the large-j asymptotics of the vertex amplitude Eq.(\ref{asymp})
\be
A_v[g_f]_{GR}\sim \sum_{p=\pm1}\sum_{j_f}N_p \exp\lt[i\sum_f j_f\lt(p\b\vartheta_f-\xi_f\rt)\rt]\ \exp\lt[-\sum_f\lt(j_f-\frac{\ca_f}{\ell^2_p\b}\rt)^2\frac{t}{2}+\sum_f\lt(\frac{\ca_f}{\ell_p^2\b}\rt)^2\frac{t}{2}\rt]
\ee
where the factor $\exp[i\sum_f j_f\lt(p\b\vartheta_f-\xi_f\rt)]$ is a rapid oscillating phase as $j_f$ is large, so that the sum over $j_f$ suppresses unless the boundary data $\xi_f$ coincide with $\b\vartheta_f$ or $-\b\vartheta_f$. Without losing generality, we assume the boundary data $\xi=\b\vartheta_f$, thus in $A_v[g_f,g_{ve}]_{GR}$ the terms with $p=+1$ are preserved in the sum, corresponding to $e^{iS_{Regge}}$, while the terms with $p=-1$ suppress in the sum over $j_f$. Therefore\footnote{It is argued in \cite{orientation,HZ} that the two different exponentials $e^{iS_{Regge}}$ and $e^{-iS_{Regge}}$ from spinfoam model corresponding to two different spacetime orientations. The fermion coupling considered here only couples to a specified spacetime orientation, which is consistent with the formulation of usual quantum field theory (in curved spacetime) \cite{hollands}.}
\be
A_v[g_f]_{GR}\sim \sum_{j_f}N_+ \exp\lt[i\sum_f j_f\lt(\b\vartheta_f-\xi_f\rt)\rt]\ \exp\lt[-\sum_f\lt(j_f-\frac{\ca_f}{\ell^2_p\b}\rt)^2\frac{t}{2}+\sum_f\lt(\frac{\ca_f}{\ell_p^2\b}\rt)^2\frac{t}{2}\rt]
\ee
If we make the PCT transformation $\Theta$ on the spin-foam amplitude $A_v[g_f,g_{ve}]_{GR}$, the reversal of the internal edge orientations doesn't affect the amplitude while the complex conjugation gives
\be
\Theta A_v[g_f]_{GR}&=&A_v[g_f]_{GR}^*\nonumber\\
&\sim& \sum_{j_f}N^*_+ \exp\lt[-i\sum_f j_f\lt(\b\vartheta_f-\xi_f\rt)\rt]\ \exp\lt[-\sum_f\lt(j_f-\frac{\ca_f}{\ell^2_p\b}\rt)^2\frac{t}{2}+\sum_f\lt(\frac{\ca_f}{\ell_p^2\b}\rt)^2\frac{t}{2}\rt]
\ee
where effectively the terms corresponding to $e^{-iS_{Regge}}$ is preserved while the terms corresponding to $e^{iS_{Regge}}$ are suppressed. The flip from $e^{iS_{Regge}}$ to $e^{-iS_{Regge}}$ corresponds to the flip of the extrinsic angles from $\vartheta_f$ to $-\vartheta_f$, which is a discrete analog of flipping the sign of the extrinsic curvature by flipping the time-orientation.

When we consider the fermions in spin-foam model, in terms of the semiclassical limit, the fermion coupling $e^{iS_F}$ should couple with the $e^{iS_{Regge}}$-terms (with positive coupling constant) on the spacetime manifold $\cm$. It means that one should specify the boundary data such that $e^{-iS_{Regge}}$-terms are suppressed, in order to reproduce the right equation of motion of graviy-fermion system on $\cm$ with a positive coupling constant $\kappa=8\pi G$. However when we flip the time and space orientation, $e^{iS_{Regge}}$-terms are suppressed while $e^{-iS_{Regge}}$-terms representing the spacetime(s) $\overline{\cm}$. Then the fermion coupling has to flip sign as $e^{-iS_F}$ to couple with $e^{-iS_{Regge}}$, in order to keep the coupling constant positive and a right equation of motion. This gives an argument\footnote{The semiclassical arguments in this paragraph and the previous paragraph are heuristic arguments supporting the definition of the PCT transformation $\Theta$. A more precise argument relies on the large-j asymptotic analysis for the gravity-fermion spin-foam model on a large number of 4-simplices, which will be studied in the future publication. The asymptotic analysis for pure gravity spinfoam on a triangulation has been done recently \cite{HZ}.} for the minus sign in the result of Eq.(\ref{-i}).

We consider a fermion correlation function of charge-conjugated fermion fields $\psi^C_v\equiv\theta\psi_v\theta^{-1}$ and $\overline{\psi}{}^C_v\equiv\theta\overline{\psi}_v\theta^{-1}$ on a PCT transformed gravity-fermion spin-foam amplitude. i.e. ( Spinfoam($\overline{\cm}$) denotes the spinfoam analog of the spacetime $\overline{\cm}$ )
\be
&&\lag\overline{\psi}{}^C_{v_1}\cdots\overline{\psi}{}^C_{v_n}\psi^C_{v_{n+1}}\cdots\psi^C_{v_{n+m}}\rag_{\text{Spinfoam}(\overline{\cm})}\nonumber\\
&:=&\sum_{j_f,i_e}\int_{\Slc}\rmd g_{ve}\prod_{f}A_f[j_f]_{GR}\prod_{v}A_v[j_f,i_e,g_{ve}]^*_{GR}\cdot f^*_{\g,j_f,i_e}\times\nonumber\\
&&\ \ \ \ \ \times\int[\cd\psi_v\cd\overline{\psi}_v]\ \overline{\psi}{}^C_{v_1}\cdots\overline{\psi}{}^C_{v_n}\psi^C_{v_{n+1}}\cdots\psi^C_{v_{n+m}}\ \exp\Big(-iS_F\lt[\ck^{-1},j_f,i_e,g_{ve},\psi_v\rt]\Big)
\ee
First of all we note that the the fermion measure is invariant under charge-conjugate
\be
[\cd\psi_v\cd\overline{\psi}_v]=[\cd\psi^{C}_v\cd\overline{\psi}{}^{C}_v]
\ee
We then make a change of variables for the fermionic integration
\be
\psi_v\mapsto\psi'_v=-i\g^5\g^0\overline{\psi}_v^\dagger=-i\g^5\psi_v\ \ \ \ \ \overline{\psi}_v\mapsto\overline{\psi}{}'_v=-i\psi_v^\dagger\g^0\g^5=-i\overline{\psi}_v\g^5
\ee
which doesn't change the integration. Then
\be
&&\lag\overline{\psi}{}^C_{v_1}\cdots\overline{\psi}{}^C_{v_n}\psi^C_{v_{n+1}}\cdots\psi^C_{v_{n+m}}\rag_{\text{Spinfoam}(\overline{\cm})}\nonumber\\
&=&\sum_{j_f,i_e}\int_{\Slc}\rmd g_{ve}\prod_{f}A_f[j_f]_{GR}\prod_{v}A_v[j_f,i_e,g_{ve}]^*_{GR}\cdot f^*_{\g,j_f,i_e}\times\nonumber\\
&&\ \ \ \ \ \times\int[\cd\psi^*_v\cd\overline{\psi}{}^*_v]\ \overline{\psi}{}^*_{v_1}\cdots\overline{\psi}{}^*_{v_n}\psi^*_{v_{n+1}}\cdots\psi^*_{v_{n+m}}\ \exp\Big(-iS_F\lt[\ck^{-1},j_f,i_e,g_{ve},\psi'_v\rt]\Big)
\ee
Again we consider the bilinear form $2iV_e\overline{\psi}{}'_v G_{ve}\g^0 G_{ev'}\psi'_{v'}$ appearing in the discretized fermion action,
\be
&&2iV_e\overline{\psi}{}'_v G_{ve}\g^0 G_{ev'}\psi_{v'}'\nonumber\\
&=& 2iV_e (-i)\psi_v^\dagger\g^0\g^5 {G}_{ve}\g^0 {G}_{ev'}(-i)\g^5\g^0\overline{\psi}_{v'}^\dagger\
=\ 2iV_e\psi_v^\dagger\g^0 {G}_{ve}\g^0 {G}_{ev'}\g^0\overline{\psi}_{v'}^\dagger\
=\ 2iV_e\psi_v^\dagger{G}^\dagger_{e v}\g^0 {G}^\dagger_{v'e}\overline{\psi}_{v'}^\dagger\nonumber\\
&=& -2iV_e\overline{\psi}{}^*_{v'} G^*_{v'e}\g^0 G^*_{ev}\psi^*_{v}
\ee
where we used the fact that $\g^5$ commutes with $G$ and the relation $\g^0{G}\g^0=({G}^{-1})^\dagger$, as well as the anticommutativity of $\psi_v$. $S_e[\psi'_{b(e)},\psi'_{f(e)},g_{ve}]$ reads
\be
S_e[\psi'_{b(e)},\psi'_{f(e)},g_{ve}]&=&-2i V_e\lt[\overline{\psi}{}^*_{f(e)}\ G^*_{f(e)\t_e}\g^0 G^*_{\t_eb(e)}\psi^*_{b(e)}-\overline{\psi}{}^*_{b(e)}G^*_{b(e)\t_e}\g^0G^*_{\t_ef(e)}\ \psi^*_{f(e)}\rt]\nonumber\\
&&=\ S_{e^{-1}}[\psi_{b(e^{-1})},\psi_{f(e^{-1})},g_{ve}]^*
\ee
which can be viewed as a nontrivial transformation property of the spinfoam-fermion action, and such a property results in the PCT symmetry of the spinfoam fermion. For the mass term
\be
S_v[\psi_v']=-m_0\fvol_v\overline{\psi}{}'_v\psi'_v=m\overline{\psi}{}^*_v\psi^*_v=S_v[\psi_v]^*
\ee
Therefore we obtain that
\be
S_F\lt[\ck^{-1},j_f,i_e,g_{ve},\psi'_v\rt]=S_F\lt[\ck,j_f,i_e,g_{ve},\psi_v\rt]^*
\ee
Then as a result
\be
&&\lag\overline{\psi}{}^C_{v_1}\cdots\overline{\psi}{}^C_{v_n}\psi^C_{v_{n+1}}\cdots\psi^C_{v_{n+m}}\rag_{\text{Spinfoam}(\overline{\cm})}\nonumber\\
&=&\sum_{j_f,i_e}\int_{\Slc}\rmd g_{ve}\prod_{f}A_f[j_f]_{GR}\prod_{v}A_v[j_f,i_e,g_{ve}]^*_{GR}\cdot f^*_{\g,j_f,i_e}\times\nonumber\\
&&\ \ \ \ \ \times\int[\cd\psi^*_v\cd\overline{\psi}{}^*_v]\ \overline{\psi}{}^*_{v_1}\cdots\overline{\psi}{}^*_{v_n}\psi^*_{v_{n+1}}\cdots\psi^*_{v_{n+m}}\ \exp\Big(iS_F\lt[\ck,j_f,i_e,g_{ve},\psi_v\rt]\Big)^*\nonumber\\
&=&\lag\overline{\psi}_{v_1}\cdots\overline{\psi}_{v_n}\psi_{v_{n+1}}\cdots\psi_{v_{n+m}}\rag^*_{\text{Spinfoam}({\cm})}
\ee
We summarize the result as a spin-foam analog of PCT theorem:

\begin{Proposition} $\mathbf{(PCT\ invariance\ for\ Spinfoam\ fermions)}$

The correlation functions of fermions on the spin-foam analog of a spacetime $\cm$ with a certain time and space orientation $o=(T,\eps_{\a\b\g\delta})$ equals the complex conjugated correlation functions of charge-conjugated fermions on the spin-foam analog of the spacetime $\overline{\cm}$ with an opposite time and space orientation $-o=(-T,\eps_{\a\b\g\delta})$, i.e.
\be
\lag\overline{\psi}{}^C_{v_1}\cdots\overline{\psi}{}^C_{v_n}\psi^C_{v_{n+1}}\cdots\psi^C_{v_{n+m}}\rag_{\text{Spinfoam}(\overline{\cm})}=\lag\overline{\psi}_{v_1}\cdots\overline{\psi}_{v_n}\psi_{v_{n+1}}\cdots\psi_{v_{n+m}}^{}\rag^*_{\text{Spinfoam}({\cm})}
\ee

\end{Proposition}

\section{Determinant of Dirac Operator on Spin-foam}\label{D}

To simplify the formula, now we consider a massless chiral Weyl fermion on the complex $\ck$
\be
S_e[\xi_{b(e)},\xi_{f(e)},g_{\t_e b(e)},g_{\t_e f(e)}]&=&2i V_e \lt[{\xi}_{b(e)}^\dagger g_{\t_e b(e)}^\dagger g_{\t_ef(e)}\xi_{f(e)}-{\xi}_{f(e)}^\dagger g_{\t_e f(e)}^\dagger g_{\t_eb(e)}\xi_{b(e)}\rt]\nonumber\\
S_F[\xi_v,g_{ve}]&=&\sum_eS_e[\psi_{b(e)},\psi_{f(e)},g_{\t_e b(e)},g_{\t_e f(e)}]\ =\ \sum_{v,v'}i\xi^\dagger_v\sD_{v,v'}\xi_v
\ee
where the Dirac operator has the following matrix element
\be
\sD^{A'A}_{b(e),f(e)}={2}V_e\ \delta_{B'B}\overline{g}_{\t_e b(e)}{}^{B'}_{\ A'} g_{\t_ef(e)}{}^B_{\ A}\ \ \ \ \ \sD_{f(e),b(e)}^{A'A}=-{2}V_e\ \delta_{B'B} \overline{g}_{\t_e f(e)}{}^{B'}_{\ A'} g_{\t_eb(e)}{}^B_{\ A}
\ee
and all the other matrix elements vanish, where we see that $\sD$ is an anti-Hermitian matrix
\be
\overline{\sD}_{v',v}=-\sD_{v,v'}
\ee
We define the Grassmann path integral measure by
\be
D\mu[\xi,\xi^\dagger]:=\prod_v\rmd(\xi_v)^1\rmd(\xi_v^\dagger)_1\rmd(\xi_v)^2\rmd(\xi_v^\dagger)_2.
\ee
Obviously the path integral of the fermion action gives the determinant of the Dirac operator (Dirac determinant)
\be
\det\sD[g_{ve}]=\int D\mu[\xi,\xi^\dagger]\ e^{iS_F[\xi_v,g_{ve}]}=\int D\mu[\xi,\xi^\dagger]\ e^{-\sum_{v,v'}\xi^\dagger_v\sD_{v,v'}\xi_v}
\ee

\subsection{Polymer representation}

\begin{figure}[h]
\begin{center}
\includegraphics[width=7cm]{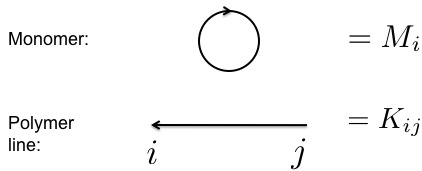}
\caption{The building blocks of polymers: a \emph{monomer} at a single index point and a \emph{polymer line} connecting two different index points}
\label{polymer}
\end{center}
\end{figure}

We briefly recall the polymer representation of the Grassmann Gauss integral \cite{lattice}. Given a matrix $Q$ such that
\be
Q_{ij}=M_i\delta_{ij}-K_{ij}
\ee
where $M_i$ are diagonal contributions and $K_{ij}$ are off-diagonal contributions, so the diagonal elements of $K_{ij}$ are assumed to vanish $K_{ii}=0$. In terms of the Grassmann Gauss integral:
\be
\det Q=\int\rmd\overline{\eta}_1\rmd\eta_1\cdots\rmd\overline{\eta}_N\rmd\eta_N\ e^{-\sum_iM_i\overline{\eta}_i\eta_i+\sum_{i\neq j}\overline{\eta}_iK_{ij}\eta_j}
\ee
This determinant can be represented in the following way: we first draw $N$ points as a lattice representing the indices $i=1,\cdots,N$. Then we draw all possible polymer diagrams using the building blocks in Fig.\ref{polymer}, such that every index point has precisely one incoming and outgoing line (a monomer is counted both as a incoming and outgoing line). We denote the set of all possible polymer diagrams by $\cp$ and denote a polymer diagram by $z$.

For each polymer diagram we write down its contribution for $\det Q$
\be
R_z=(-1)^{\text{Number of Polymer Loops}}\prod_iM_i\prod_{j,k}K_{jk}
\ee
where the index point $i$ is attached by a monomer $M_i$ and the index points $j$ and $k$ are connected by a polymer line $K_{jk}$, while the polymer lines forming closed polymer loops. Then the determinant $\det Q$ equals the sum over all possible polymer contributions:
\be
\det Q=\sum_{z\in\cp}R_z
\ee

Now we consider the determinant of our Dirac operator $\sD$. The diagonal elements of $\sD$ are zeros, thus the monomer contribution is not allowed. For the off-diagonal elements:
\be
\sD_{(b(e),A');(f(e),A)}={2}V_e\ \delta_{B'B}(\overline{g}_{e b(e)})_{\ A'}^{B'} (g_{ef(e)})^B_{\ A}\ \ \ \ \ \sD_{(f(e),A');(b(e),A)}=-{2}V_e\ \delta_{B'B} (\overline{g}_{e f(e)})_{\ A'}^{B'}(g_{eb(e)})^B_{\ A}
\ee
We draw the lattice of all the index points $j\equiv(v,A)$, which is $V(\ck)\times \mathbb{Z}_2$. Then we draw all possible polymer loop diagrams using polymer lines $\sD_{ij}$, such that every index point has precisely one incoming and outgoing line. For each polymer diagram $z\in\cp$
\be
R_z&=&(-1)^{\text{Number of Polymer Loops}}\prod_{j,k}(-\sD_{jk})\nonumber\\
&=&(-1)^{\text{Number of Polymer Loops}}\prod_{(v',A');(v,A)}(-1)^{\eps_{v'v}}(2V_e)\delta_{B'B} (\overline{g}_{e v'})_{\ A'}^{B'}(g_{ev})^B_{\ A}
\ee
where $\eps_{v'v}=1$ if $v'=b(e)$ and $v=f(e)$ otherwise $\eps_{v'v}=0$. Finally the Dirac determinant is represented as a sum over all polymer contributions:
\be
\det\sD=\sum_{z\in\cp}R_z
\ee

For the massive Dirac fermion, the lattice of all the index points is $V(\ck)\times\{1,2,3,4\}$. And we not only need to consider the polymer lines, but also need to consider the monomers $iM_{j=(v,\a)}=im_0\fvol_v$. For each polymer diagram $z\in\cp$ we have
\be
R_z&=&(-1)^{\text{Number of Polymer Loops}}\prod_liM_l\prod_{j,k}(-\sD_{jk})\nonumber\\
&=&(-1)^{\text{Number of Polymer Loops}}\prod_{(v,\a)}\big(im_0\fvol_v\big)\prod_{(v',\a');(v,\a)}(-1)^{\eps_{v'v}}\big(2V_e\big)\big[G_{v'e}\g^0G_{ev}\big]_{\a'\a}
\ee
And
\be
\det(\sD+iM)=\sum_{z\in\cp}R_z
\ee

\subsection{$\eps$-loop representation}

Here we compute the Grassmann path integral in a more explicite manner. Because of the Grassmann variables, on each edge the exponentiated fermion action has a 9-term expansion
\be
e^{iS_e}
&=&\lt[1-{\xi}_{b(e)}^\dagger \sD_{b(e),f(e)}\xi_{f(e)}+\half\lt({\xi}_{b(e)}^\dagger \sD_{b(e),f(e)}\xi_{f(e)}\rt)^2\rt]\times\nonumber\\
&&\lt[1-{\xi}_{f(e)}^\dagger \sD_{f(e),b(e)}\xi_{b(e)}+\half\lt({\xi}_{f(e)}^\dagger \sD_{f(e),b(e)}\xi_{b(e)}\rt)^2\rt]\nonumber\\
&=&1-{\xi}_{b(e)}^\dagger \sD_{b(e),f(e)}\xi_{f(e)}-{\xi}_{f(e)}^\dagger \sD_{f(e),b(e)}\xi_{b(e)}+\lt({\xi}_{b(e)}^\dagger \sD_{b(e),f(e)}\xi_{f(e)}\rt)\lt({\xi}_{f(e)}^\dagger \sD_{f(e),b(e)}\xi_{b(e)}\rt)\nonumber\\
&&+\half\lt({\xi}_{b(e)}^\dagger \sD_{b(e),f(e)}\xi_{f(e)}\rt)^2+\half\lt({\xi}_{f(e)}^\dagger \sD_{f(e),b(e)}\xi_{b(e)}\rt)^2\nonumber\\
&&-\half\lt({\xi}_{b(e)}^\dagger \sD_{b(e),f(e)}\xi_{f(e)}\rt)\lt({\xi}_{f(e)}^\dagger \sD_{f(e),b(e)}\xi_{b(e)}\rt)^2
-\half\lt({\xi}_{f(e)}^\dagger \sD_{f(e),b(e)}\xi_{b(e)}\rt)\lt({\xi}_{b(e)}^\dagger \sD_{b(e),f(e)}\xi_{f(e)}\rt)^2\nonumber\\
&&+\frac{1}{4}\lt({\xi}_{b(e)}^\dagger \sD_{b(e),f(e)}\xi_{f(e)}\rt)^2\lt({\xi}_{f(e)}^\dagger \sD_{f(e),b(e)}\xi_{b(e)}\rt)^2\label{edgeaction}
\ee
Given a Grassmann integration at a vertex $v$, it only affects the $e^{iS_e}$'s with the $e$'s connecting to $v$. We have
\be
\int\rmd(\xi_v)^1\rmd(\xi_v^\dagger)_1\rmd(\xi_v)^2\rmd(\xi_v^\dagger)_2\prod_{e,b(e)=v}e^{iS_e}\prod_{e',f(e')=v}e^{iS_{e'}}
\ee
We know that the only nonvanishing Grassmann integral is
\be
\int\rmd\xi\rmd\xi^\dagger\ \overline{\xi}{}^{A'}\overline{\xi}{}^{B'}{\xi}^{A}{\xi}^{B}\equiv\int\rmd\xi^1\rmd(\xi^\dagger)_1\rmd\xi^2\rmd(\xi^\dagger)_2\ \overline{\xi}{}^{A'}\overline{\xi}{}^{B'}{\xi}^{A}{\xi}^{B}=-\eps^{A'B'}\eps^{AB}
\ee
Therefore there are a few examples for the possible contributions for the integral at each vertex $v$

\begin{description}

\item[Example 1.] Consider two outgoing edges $e_1,e_2$ and two incoming edges $e_3,e_4$, $v=b(e_1)=b(e_2)=f(e_3)=f(e_4)$, we have a nonvanishing integral
\be
&&\int\rmd\xi_v\rmd\xi_v^\dagger\lt[\overline{\xi}_{v}^{A'} \sD_{v,f(e_1)}{}_{A' A}\xi_{f(e_1)}^A\ \overline{\xi}_{f(e_2)}^{B'} \sD_{f(e_2),v}{}_{B' B}\xi_v^B\rt]\lt[\overline{\xi}_{b(e_3)}^{C'} \sD_{b(e_3),v}{}_{C' C}\xi_{v}^C\ \overline{\xi}_{v}{}^{D'} \sD_{v,b(e_4)}{}_{D' D}\xi_{b(e_4)}^D\rt]\nonumber\\
&=&\int\rmd\xi_v\rmd\xi_v^\dagger\ \ \overline{\xi}_{v}^{A'}\xi_v^B\xi_{v}^C\ \overline{\xi}_{v}^{D'} \lt[\sD_{v,f(e_1)}{}_{A' A}\xi_{f(e_1)}^A\ \overline{\xi}_{f(e_2)}^{B'} \sD_{f(e_2),v}{}_{B' B}\rt]\lt[\overline{\xi}_{b(e_3)}^{C'} \sD_{b(e_3),v}{}_{C' C}\sD_{v,b(e_4)}{}_{D' D}\xi_{b(e_4)}^D\rt]\nonumber\\
&=&-\eps^{A'D'}\eps^{BC} \lt[\sD_{v,f(e_1)}{}_{A' A}\xi_{f(e_1)}^A\ \overline{\xi}_{f(e_2)}^{B'} \sD_{f(e_2),v}{}_{B' B}\rt]\lt[\overline{\xi}_{b(e_3)}^{C'} \sD_{b(e_3),v}{}_{C' C}\sD_{v,b(e_4)}{}_{D' D}\xi_{b(e_4)}^D\rt]\nonumber\\
&=&-\lt(\overline{\xi}_{f(e_2)}^{B'}\overline{\xi}_{b(e_3)}^{C'} \sD_{f(e_2),v}{}_{B' B} \sD_{b(e_3),v}{}_{C' C}\eps^{BC}\rt)\lt(\eps^{A'D'}\sD_{v,f(e_1)}{}_{A' A}\sD_{v,b(e_4)}{}_{D' D}\xi_{f(e_1)}^A\xi_{b(e_4)}^D\rt)
\ee
Insert in the matrix element of Dirac operator
\be
\sD_{b(e),f(e)}{}_{A'A}={2}V_e\ \delta_{B'B}(\overline{g}_{e b(e)})_{\ A'}^{B'} (g_{ef(e)})^B_{\ A}\ \ \ \ \ \sD_{f(e),b(e)}{}_{A'A}=-{2}V_e\ \delta_{B'B} (\overline{g}_{e f(e)})_{\ A'}^{B'}(g_{eb(e)})^B_{\ A}
\ee
the above integral equals
\be
&=&-16{V_{e_1}V_{e_2}V_{e_3}V_{e_4}}\lt[\overline{\xi}_{f(e_2)}^{B'}\overline{\xi}_{b(e_3)}^{C'}\overline{g}_{e_2 f(e_2)}{}_{\ B'}^{E'}\overline{g}_{e_3 b(e_3)}{}_{\ C'}^{F'}\delta_{E' E}\delta_{F' F} g_{e_2 v}{}^E_{\ B} g_{e_3v}{}^F_{\ C}\eps^{BC}\rt]\nonumber\\
&&\ \ \ \ \ \times\ \lt[\eps^{A'D'}\overline{g}_{e_1 v}{}_{\ A'}^{G'}\overline{g}_{e_4 v}{}_{\ D'}^{H'}\delta_{G' G}\delta_{H' H}g_{e_1f(e_1)}{}^G_{\ A} g_{e_4 b(e_4)}{}^{H}_{\ D}\xi_{f(e_1)}^A\xi_{b(e_4)}^D\rt]
\ee

\item[Example 2.] Consider two outgoing edges $e_1,e_2$, $b(e_1)=b(e_2)=v$, there is a nonvanishing integral
\be
&&\int\rmd\xi_v\rmd\xi_v^\dagger\ \frac{1}{2}\lt[\overline{\xi}_{v}{}^{A'} \sD_{v,f(e_1)}{}_{A' A}\xi_{f(e_1)}^A\ \overline{\xi}_{v}{}^{B'} \sD_{v,f(e_1)}{}_{B' B}\xi^B_{f(e_1)}\rt]\lt[\overline{\xi}_{f(e_1)}{}^{C'} \sD_{f(e_1),v}{}_{C' C}\xi^C_{v}\ \overline{\xi}_{f(e_2)}{}^{D'}\sD_{f(e_2),v}{}_{D' D}\xi_{v}^D\rt]\nonumber\\
&=&\int\rmd\xi_v\rmd\xi_v^\dagger\ \frac{1}{2}\lt[\overline{\xi}_{v}{}^{A'}\overline{\xi}_{v}{}^{B'} \sD_{v,f(e_1)}{}_{A' A}\sD_{v,f(e_1)}{}_{B' B}\xi_{f(e_1)}^A\xi^B_{f(e_1)}\rt]\lt[\overline{\xi}_{f(e_1)}{}^{C'}\overline{\xi}_{f(e_2)}{}^{D'} \sD_{f(e_1),v}{}_{C' C}\sD_{f(e_2),v}{}_{D' D}\xi^C_{v}\xi_{v}^D\rt]\nonumber\\
&=&-\frac{1}{2}\lt[\eps^{A'B'} \sD_{v,f(e_1)}{}_{A' A}\sD_{v,f(e_1)}{}_{B' B}\xi_{f(e_1)}^A\xi^B_{f(e_1)}\rt]\lt[\overline{\xi}_{f(e_1)}{}^{C'}\overline{\xi}_{f(e_2)}{}^{D'} \sD_{f(e_1),v}{}_{C' C}\sD_{f(e_2),v}{}_{D' D}\eps^{CD}\rt]\nonumber\\
&=&-8{V_{e_1}^3V_{e_2}}\lt[\eps^{A'B'} (g_{e_1 v}^\dagger g_{e_1f(e_1)})_{A' A}(g_{e_1 v}^\dagger g_{e_1f(e_1)})_{B' B}\xi_{f(e_1)}^A\xi^B_{f(e_1)}\rt]\nonumber\\
&&\times\lt[\overline{\xi}{}_{f(e_1)}^{C'}\overline{\xi}{}_{f(e_2)}^{D'} (g_{e_1 f(e_1)}^\dagger g_{e_1 v})_{C' C}(g_{e_2 f(e_2)}^\dagger g_{e_2 v})_{D' D}\eps^{CD}\rt]\nonumber\\
&=&-8{V_{e_1}^3V_{e_2}}\lt[\eps_{AB}\xi_{f(e)}^A\xi^B_{f(e)}\rt]\lt[\overline{\xi}{}_{f(e_1)}^{C'}\overline{\xi}{}_{f(e_2)}^{D'} (g_{e_1 f(e_1)}^\dagger g_{e_1 v})_{C' C}(g_{e_2 f(e_2)}^\dagger g_{e_2 v})_{D' D}\eps^{CD}\rt]
\ee

\item[Example 3.] Consider an outgoing edges $e_1$, and an incoming edge $e_2$, $b(e_1)=f(e_2)=v$, there is a nonvanishing integral
\be
&&\int\rmd\xi_v\rmd\xi_v^\dagger\ \frac{1}{2}\lt[\overline{\xi}_{v}{}^{A'} \sD_{v,f(e_1)}{}_{A' A}\xi_{f(e_1)}^A\ \overline{\xi}_{v}{}^{B'} \sD_{v,f(e_1)}{}_{B' B}\xi^B_{f(e_1)}\rt]\lt[\overline{\xi}_{f(e_1)}{}^{C'} \sD_{f(e_1),v}{}_{C' C}\xi^C_{v}\ \overline{\xi}_{b(e_2)}{}^{D'}\sD_{f(e_2),v}{}_{D' D}\xi_{v}^D\rt]\nonumber\\
&=&-\frac{1}{2}\lt[\eps^{A'B'} \sD_{v,f(e_1)}{}_{A' A}\sD_{v,f(e_1)}{}_{B' B}\xi_{f(e_1)}^A\xi^B_{f(e_1)}\rt]\lt[\overline{\xi}_{f(e_1)}{}^{C'}\overline{\xi}_{f(e_2)}{}^{D'} \sD_{f(e_1),v}{}_{C' C}\sD_{b(e_2),v}{}_{D' D}\eps^{CD}\rt]\nonumber\\
&=&-8V_{e_1}^3V_{e_2}\lt[\eps^{A'B'} (g_{e_1 v}^\dagger g_{e_1f(e_1)})_{A' A}(g_{e_1 v}^\dagger g_{e_1f(e_1)})_{B' B}\xi_{f(e_1)}^A\xi^B_{f(e_1)}\rt]\nonumber\\
&&\times(-1)\lt[\overline{\xi}{}_{f(e_1)}^{C'}\overline{\xi}{}_{f(e_2)}^{D'} (g_{e_1 f(e_1)}^\dagger g_{e_1 v})_{C' C}(g_{e_2 b(e_2)}^\dagger g_{e_2 v})_{D' D}\eps^{CD}\rt]\nonumber\\
&=&-8{V_{e_1}^3V_{e_2}}\lt[\eps_{AB}\xi_{f(e)}^A\xi^B_{f(e)}\rt](-1)\lt[\overline{\xi}{}_{f(e_1)}^{C'}\overline{\xi}{}_{f(e_2)}^{D'} (g_{e_1 f(e_1)}^\dagger g_{e_1 v})_{C' C}(g_{e_2 b(e_2)}^\dagger g_{e_2 v})_{D' D}\eps^{CD}\rt]
\ee

\item[Example 4.] For each single outgoing edge $e$, $v=b(e)$, we have the integral
\be
&&\int\rmd\xi_v\rmd\xi_v^\dagger\ \frac{1}{4}\lt[\overline{\xi}_{v}{}^{A'} \sD_{v,f(e)}{}_{A' A}\xi_{f(e)}^A\ \overline{\xi}_{v}{}^{B'} \sD_{v,f(e)}{}_{B' B}\xi^B_{f(e)}\rt]\lt[\overline{\xi}_{f(e)}{}^{C'} \sD_{f(e),v}{}_{C' C}\xi^C_{v}\ \overline{\xi}_{f(e)}{}^{D'}\sD_{f(e),v}{}_{D' D}\xi_{v}^D\rt]\nonumber\\
&=&-4{V_e^4}\lt[\eps_{AB}\xi_{f(e)}^A\xi^B_{f(e)}\rt]\lt[\overline{\xi}{}_{f(e)}^{C'}\overline{\xi}{}_{f(e)}^{D'} \eps_{C'D'}\rt]
\ee

\item[Example 5.] For each single incoming edge $e$, $v=f(e)$, in the same way
\be
&&\int\rmd\xi_v\rmd\xi_v^\dagger\ \frac{1}{4}\lt[\overline{\xi}_{v}{}^{A'} \sD_{v,b(e)}{}_{A' A}\xi_{f(e)}^A\ \overline{\xi}_{v}{}^{B'} \sD_{v,b(e)}{}_{B' B}\xi^B_{b(e)}\rt]\lt[\overline{\xi}_{b(e)}{}^{C'} \sD_{b(e),v}{}_{C' C}\xi^C_{v}\ \overline{\xi}_{b(e)}{}^{D'}\sD_{b(e),v}{}_{D' D}\xi_{v}^D\rt]\nonumber\\
&=&-4{V_e^4}\lt[\eps_{AB}\xi_{b(e)}^A\xi^B_{b(e)}\rt]\lt[\overline{\xi}{}_{b(e)}^{C'}\overline{\xi}{}_{b(e)}^{D'} \eps_{C'D'}\rt]
\ee

\end{description}

All the contributions of the determinant $\det\sD$ can be obtained by the integrals for each vertex, similar to the previous examples. And they can be represented graphically:

\begin{itemize}
\item We draw an arrow-line for each $\xi^\dagger_v\sD_{v,v'}\xi_v'$ in $S_e$ and associate $\xi^\dagger$ to its source and $\xi$ to its target. The arrow-line (Fig.\ref{line}) represents $\sD_{v,v'}{}_{A'A}$ with $A'$ at its source and $A$ at its target (however in the follow graphic representation we often ignore the $A',A$ label for the arrow-line, in order to simplify the graph).

\begin{figure}[h]
\begin{center}
\includegraphics[width=5cm]{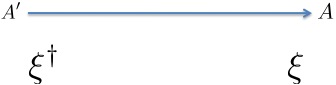}
\caption{A fermion arrow-line.}
\label{line}
\end{center}
\end{figure}

\item For each edge we represent $S_e$ by Fig.\ref{edge}, where we have a minus sign $(-1)$ when orientation of a fermion arrow line in $S_e$ coincides with the orientation of the corresponding edge $e$. The parallel double-arrows correspond to the terms $(\xi^\dagger\sD\xi)^2$ in the expression of $S_e$ Eq.(\ref{edgeaction}).

\begin{figure}[h]
\begin{center}
\includegraphics[width=14cm]{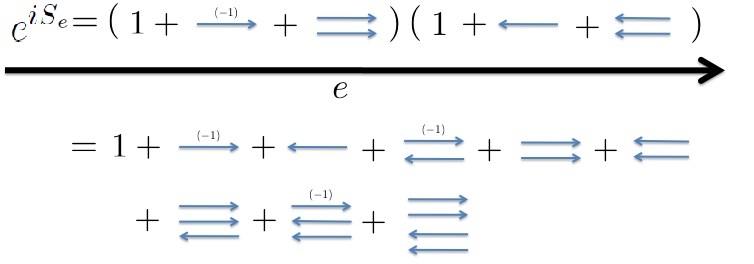}
\caption{All possible contributions from an edge, see Eq.(\ref{edgeaction}) for the correspondence.}
\label{edge}
\end{center}
\end{figure}

\item We assign the weights to the arrows and double-arrows (Fig.\ref{weight})

\begin{figure}[h]
\begin{center}
\includegraphics[width=3cm]{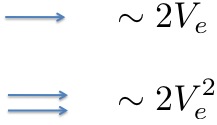}
\caption{The weights for arrow and double-arrow.}
\label{weight}
\end{center}
\end{figure}

\item We represent $\eps^{AB}$ (resp. $\eps^{A'B'}$) by a contractors connecting two targets (resp. two sources) of two arrow-lines (Fig.\ref{contractor}). Note that the $\eps$-contractors not only can contract the arrow-lines from different edges, but also can contract the double-arrows from a single edge.

\begin{figure}[h]
\begin{center}
\includegraphics[width=12cm]{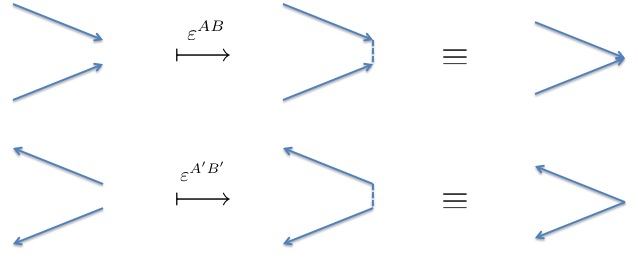}
\caption{A $\eps$-contractor connects two sources or two targets.}
\label{contractor}
\end{center}
\end{figure}

\item Each vertex $v\in V(\ck)$ must choose precisely two incoming and two outgoing fermion arrow-lines from the edges connecting $v$. These fermion arrow-lines are contracted by two $\eps$-contractors. There are 7 types of contractions shown in Fig.\ref{vertex}. Note that it requires precisely the same number of incoming and outgoing arrows because one need precisely two $\xi$ and two $\overline{\xi}$ to make the integral nonvanish.

\begin{figure}[h]
\begin{center}
\includegraphics[width=12cm]{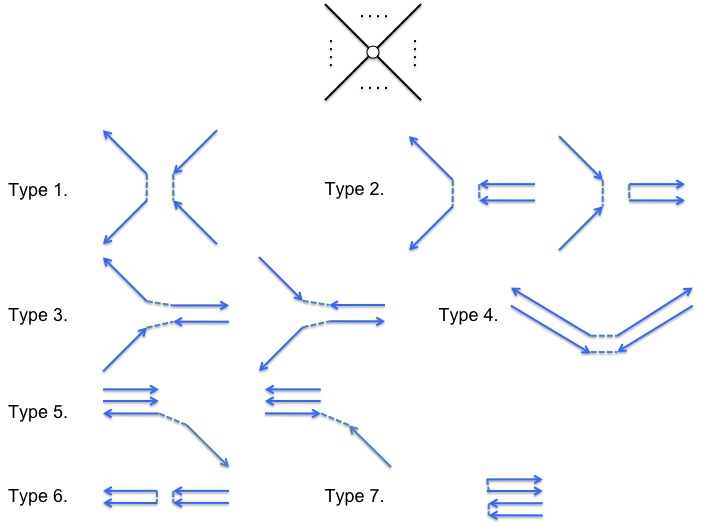}
\caption{The typical contractions for each vertex.}
\label{vertex}
\end{center}
\end{figure}

\begin{description}
\item[Type 1:] Four fermion arrow-lines are from four different edges;

\item[Type 2:] Two arrow-lines are from two different edges, but a double arrow is from another single edge;

\item[Type 3:] A pair of opposite oriented arrows from a single edge, and two arrows from two different edges;

\item[Type 4:] Two pairs of opposite oriented arrows from two different edges;

\item[Type 5:] A double-arrow and a single arrow from a single edge, and another arrow from another edge;

\item[Type 6:] two double-arrows are from two different edges;

\item[Type 7:] two double-arrows are from the same edge.
\end{description}

\item The contraction at each vertex makes the arrow-lines form close loops. There are two type of loops, i.e. nontrivial loops and trivial loops, see Fig.\ref{loop}. We call these loops the ``$\eps$-loops'' because the neighboring edges have opposite directions. Note that a double-arrow from a single edge only can form a trivial $\eps$-loop by the above contraction rule. And each $\eps$-loop must contain even number of arrow-lines by construction.

\begin{figure}[h]
\begin{center}
\includegraphics[width=10cm]{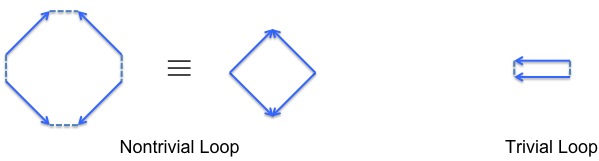}
\caption{A 4-gon nontrivial $\eps$-loop and a trivial $\eps$-loop.}
\label{loop}
\end{center}
\end{figure}

\item We denote a n-gon nontrivial $\eps$-loop by $L_n$ ($n\geq2$ is even) and a trivial $\eps$-loop by $T$. Each trivial $\eps$-loop contributes
\be
T=-2{V_e^2}\eps_{AB}\eps^{AB}=-4{V_e^2}
\ee
Let's consider a non-trivial $\eps$-loop, for example a 4-gon $\eps$-loop with vertices $v_1,v_2,v_3,v_4$ (cyclic ordered) and $v_1$ is a source node for two arrows
\be
L_4&=&\int\rmd\xi_1^\dagger\rmd\xi_2\rmd\xi_3^\dagger\rmd\xi_4\lt(\overline{\xi}{}_1^{A'}\sD_{12}{}_{A'A}\xi_2^A\rt)\lt(\overline{\xi}{}_3^{B'}\sD_{32}{}_{B'B}\xi_2^B\rt)\lt(\overline{\xi}{}_3^{C'}\sD_{34}{}_{C'C}\xi_4^C\rt)\lt(\overline{\xi}{}_1^{D'}\sD_{14}{}_{D'D}\xi_4^D\rt)\nonumber\\
&=&\eps^{D'A'}\sD_{12}{}_{A'A}(-\eps^{AB})\sD_{32}{}_{B'B}\eps^{B'C'}\sD_{34}{}_{C'C}\eps^{CD}\sD_{14}{}_{D'D}\nonumber\\
&=&(-1)16{V_{e_1}V_{e_2}V_{e_3}V_{e_4}}(-1)^{\eps_{L_4}}\times
\eps^{D'A'}\lt[\delta_{E'E}(\overline{g}_{e_1 v_1})_{\ A'}^{E'} (g_{e_1v_2})^E_{\ A}\rt]
\eps^{AB}\lt[\delta_{F'F}(\overline{g}_{e_2 v_3})_{\ B'}^{F'} (g_{e_2 v_2})^F_{\ B}\rt]\times\nonumber\\
&&\eps^{B'C'}\lt[\delta_{G'G}(\overline{g}_{e_3 v_3})_{\ C'}^{G'} (g_{e_3 v_4})^G_{\ C}\rt]
\eps^{CD}\lt[\delta_{H'H}(\overline{g}_{e_4 v_1})_{\ D'}^{H'} (g_{e_4v_4})^H_{\ D}\rt]\nonumber\\
&=&(-1)16{V_{e_1}V_{e_2}V_{e_3}V_{e_4}}(-1)^{\eps_{L_4}}\tr\lt[(\overline{g}_{e_4v_1}\eps \overline{g}^t_{e_1v_1})({g}_{e_1v_2}\eps {g}^t_{e_2v_2})( \overline{g}_{e_2v_3}\eps \overline{g}^t_{e_3v_3})( {g}_{e_3v_4}\eps {g}^t_{e_4v_4})\rt]
\ee
where $\eps_{L_n}$ is the number of arrows whose orientations coincide to the orientations of the associated edges.

\begin{figure}[h]
\begin{center}
\includegraphics[width=5cm]{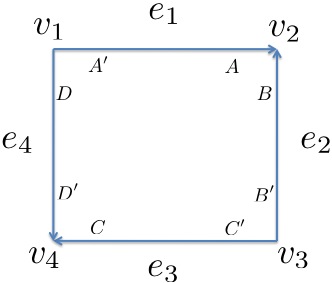}
\caption{A 4-gon nontrivial $\eps$-loop.}
\label{loop}
\end{center}
\end{figure}

Generalize it to n-gon $\eps$-loop $L_n$ (choose $v_1$ to be a source node, $v_i,e_i$ are cyclic ordered)
\be
L_n=(-1)\prod_{i=1}^n2{V_{e_i}}(-1)^{\eps_{L_n}}\tr\lt[(\overline{g}_{e_nv_1}\eps \overline{g}^t_{e_1v_1})({g}_{e_1v_2}\eps {g}^t_{e_2v_2})\cdots( \overline{g}_{e_{n-2}v_{n-1}}\eps \overline{g}^t_{e_{n-1}v_{n-1}})( {g}_{e_{n-1}v_n}\eps {g}^t_{e_nv_n})\rt]
\ee
where $g_{e,e'}$ is the $\Slc$ holonomy from the middle point $\t_e$ to the middle point $\t_{e'}$.

\item We draw all possible close $\eps$-loop diagrams $\{\G_F\}$ by using the possible edge contributions (Fig.\ref{edge}) and the possible vertex contributions (Fig.\ref{vertex}). For each $\eps$-loop diagram, it consists a certain number of trivial $\eps$-loops $T(i)$, $i=1,\cdots,|\ct_{\G_F}|$, and a certain number of nontrivial $\eps$-loops $L_n(i)$, $i=1,\cdots,|\cl_{\G_F}|$, where $\ct_{\G_F}$ and $\cl_{\G_F}$ are respectively the sets of all trivial and nontrivial loops in $\G_F$. Therefore we can write the determinant of the Dirac operator is a sum over all possible loop diagrams
\be
\det\sD&=&\sum_{\{\G_F\}}\prod_{i=1}^{|\ct_{\G_F}|}T(i)\prod_{i=1}^{|\cl_{\G_F}|}L_n(i)\label{det}\\
&=&\sum_{\{\G_F\}}(-1)^{\text{Number of Loops}}\ 4^{\text{Number of Trivial Loops}}\prod_{\{\to\}}V_e\times\nonumber\\
&&\prod_{\{\text{Nontrivial Loops}\}}\!\!\!\!\!\!(-1)^{\eps_{L_n}}\tr\lt[(\overline{g}_{e_nv_1}\eps \overline{g}^t_{e_1v_1})({g}_{e_1v_2}\eps {g}^t_{e_2v_2})\cdots( \overline{g}_{e_{n-2}v_{n-1}}\eps \overline{g}^t_{e_{n-1}v_{n-1}})( {g}_{e_{n-1}v_n}\eps {g}^t_{e_nv_n})\rt]\nonumber
\ee
where $\{\to\}$ denote the set of all the arrows in the $\eps$-loop diagram $\G_F$, so $\prod_{\{\to\}}V_e$ means the product of all the $V_e$ associated with the edges corresponding to all the arrows, and note that for an edge there can be two arrows.

\end{itemize}

One could follow the following two steps to construct each term in the sum Eq.(\ref{det})

\begin{description}
\item[Step 1.] We first ignore all the trivial $\eps$-loops. All nontrivial $\eps$-loops can be constructed by using the first 4 terms of $S_e$ in Fig.\ref{edge} and the type 1-5 vertex contributions in Fig.\ref{vertex} (ignoring trivial loops). But one should make sure that each vertex has 4 fermion arrows (2 incoming and 2 outgoing arrows) or has 2 fermion arrows (both incoming or both outgoing).

\begin{figure}[h]
\begin{center}
\includegraphics[width=5cm]{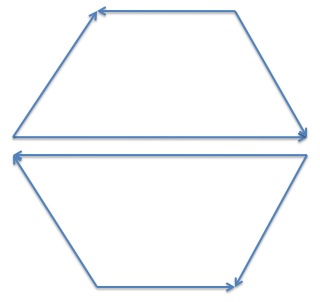}
\caption{A simple diagram with 2 nontrivial loops.}
\label{2loop}
\end{center}
\end{figure}

\item[Step 2.] We add trivial $\eps$-loops such that all the vertex has precisely 4 fermion arrows, two of which are incoming and two of which are outgoing.

\end{description}

\section{Fermion Correlation Functions on Spin-foam}\label{corr}

\subsection{Computing Fermion n-point Functions}

Now we consider the correlation functions of a massless Weyl fermion on spin-foam
\be
&&\lag{\xi}^\dagger_{v_1}\cdots{\xi}^\dagger_{v_n}\xi_{v_{n+1}}\cdots\xi_{v_{n+m}}\rag_{\text{Spinfoam}}\nonumber\\
&:=&\sum_{j_f,i_e}\int_{\Slc}\rmd g_{ve}\prod_{f}A_f[j_f]_{GR}\prod_{v}A_v[j_f,i_e,g_{ve}]_{GR}\cdot f_{\g,j_f,i_e}\times\prod_{(v,e_0)}\delta_{\Slc}(g_{ve_0})\times\nonumber\\
&&\ \ \ \ \ \times\int D\mu[\xi,\xi^\dagger]\ {\xi}^\dagger_{v_1}\cdots{\xi}^\dagger_{v_n}\xi_{v_{n+1}}\cdots\xi_{v_{n+m}}\ \exp\Big(iS_F\lt[\ck,j_f,i_e,g_{ve},\xi_v\rt]\Big)\label{correlator2}
\ee
The Weyl fermion action reads
\be
S_e[\xi_{b(e)},\xi_{f(e)},g_{\t_e b(e)},g_{\t_e f(e)}]&=&2i V_e \lt[{\xi}_{b(e)}^\dagger g_{\t_e b(e)}^\dagger g_{\t_ef(e)}\xi_{f(e)}-{\xi}_{f(e)}^\dagger g_{\t_e f(e)}^\dagger g_{\t_eb(e)}\xi_{b(e)}\rt]\nonumber\\
iS_F[\xi_v,g_{ve}]&=&\sum_eiS_e[\psi_{b(e)},\psi_{f(e)},g_{\t_e b(e)},g_{\t_e f(e)}]\ =\ -\sum_{v,v'}\xi^\dagger_v\sD_{v,v'}[j_f,i_e,g_{ve}]\xi_v
\ee
where the Dirac operator has the following matrix element
\be
\sD_{b(e),f(e)}={2}V_e\ g_{\t_e b(e)}^\dagger g_{\t_ef(e)}\ \ \ \ \ \sD_{f(e),b(e)}=-{2}V_e\ g_{\t_e f(e)}^\dagger g_{\t_e b(e)}
\ee
We employ the standard textbook technique to evaluate the correlation functions. We define a generating functional
\be
Z(\ck,\eta^\dagger,{\eta})_f&:=&\sum_{j_f,i_e}\int_{\Slc}\rmd g_{ve}\prod_{(v,e_0)}\delta_{\Slc}(g_{ve_0})\prod_{f}A_f[j_f]_{GR}\prod_{v}A_v[j_f,i_e,g_{ve}]_{GR}\cdot f_{\g,j_f,i_e}\times\nonumber\\
&&\ \ \ \ \ \times\int D\mu[\xi,\xi^\dagger]\ \exp\Big(-\sum_{v,v'}\xi^\dagger_v\ \sD_{v,v'}[j_f,i_e,g_{ve}]\ \xi_v+\sum_v{\eta}^\dagger_v\xi_v+\sum_v\xi_v^\dagger\eta_v\Big)\nonumber\\
&=&\sum_{j_f,i_e}\int_{\Slc}\rmd g_{ve}\prod_{(v,e_0)}\delta_{\Slc}(g_{ve_0})\prod_{f}A_f[j_f]_{GR}\prod_{v}A_v[j_f,i_e,g_{ve}]_{GR}\cdot f_{\g,j_f,i_e}\times\nonumber\\
&&\ \ \ \ \ \times\ \det\sD[j_f,i_e,g_{ve}]\ \exp\Big(\sum_{v,v'}\eta^\dagger_v\ \sD^{-1}_{v,v'}[j_f,i_e,g_{ve}]\ \eta_v\Big)
\ee
Then the correlation function is given by the functional derivative, e.g. the 2-point function (fermion propagator) is given by
\be
&&\lag\xi_{v_1}\xi^\dagger_{v_2}\rag_{\text{Spinfoam}}\ =\ \frac{\delta^2}{\delta\eta_{v_2}\delta\eta^\dagger_{v_1}}Z(\ck,\eta^\dagger,{\eta})_f\Big|_{\eta={\eta}^\dagger=0}\nonumber\\
&&=\ \sum_{j_f,i_e}\int_{\Slc}\rmd g_{ve}\prod_{(v,e_0)}\delta_{\Slc}(g_{ve_0})\prod_{f}A_f[j_f]_{GR}\prod_{v}A_v[j_f,i_e,g_{ve}]_{GR}\cdot f_{\g,j_f,i_e}\times\nonumber\\
&&\ \ \ \ \ \times\ \sD^{-1}_{v_1,v_2}[j_f,i_e,g_{ve}]\ \det\sD[j_f,i_e,g_{ve}]
\ee
A n-point correlation function is given by
\be
&&\lag{\xi}_{v_1}\cdots{\xi}_{v_n}\xi^\dagger_{v_{n+1}}\cdots\xi^\dagger_{v_{n+m}}\rag_{\text{Spinfoam}}
\ =\ 0\ \ \ \ \ \text{if}\ \ \ m\neq n\nonumber\\
&&\lag{\xi}_{v_1}\cdots{\xi}_{v_n}\xi^\dagger_{v_{n+1}}\cdots\xi^\dagger_{v_{2n}}\rag_{\text{Spinfoam}}
\ =\ \sum_{\sig}(-1)^\sig\frac{\delta^2}{\delta\eta_{v_{n+\sig(1)}}\delta\eta^\dagger_{v_1}}\cdots \frac{\delta^2}{\delta\eta_{v_{n+\sig(n)}}\delta\eta^\dagger_{v_1}}Z(\ck,\eta^\dagger,{\eta})_f\Big|_{\eta={\eta}^\dagger=0}\nonumber\\
&&=\ \sum_{j_f,i_e}\int_{\Slc}\rmd g_{ve}\prod_{(v,e_0)}\delta_{\Slc}(g_{ve_0})\prod_{f}A_f[j_f]_{GR}\prod_{v}A_v[j_f,i_e,g_{ve}]_{GR}\cdot f_{\g,j_f,i_e}\times\nonumber\\
&&\ \ \ \ \ \times\ \det\sD[j_f,i_e,g_{ve}]\sum_{\sig}(-1)^\sig\sD^{-1}_{v_1,v_{n+\sig(1)}}[j_f,i_e,g_{ve}]\cdots\sD^{-1}_{v_n,v_{n+\sig(n)}}[j_f,i_e,g_{ve}]\label{correlator3}
\ee
where $\sig$ denotes the permutation $(1,2,\cdots,n)\mapsto(\sig(1),\sig(2),\cdots,\sig(n))$. The above result shows that the fermion correlation function on spin-foams is given by the sum over (disconnected) Feynman diagrams (for free fermion field, because here we only consider the interaction between fermions and gravity)
\be
\det\sD[j_f,i_e,g_{ve}]\sum_{\sig}(-1)^\sig\sD^{-1}_{v_1,v_{n+\sig(1)}}[j_f,i_e,g_{ve}]\cdots\sD^{-1}_{v_n,v_{n+\sig(n)}}[j_f,i_e,g_{ve}]
\ee
on the spin-foam, and the amplitude given by the Feynman diagram depends on the gravitational degree of freedom $j_f,i_e,g_{ve}$, which should also be summed over in the spin-foam model. This result is quite similar to the idea in the earlier work \cite{FL} about coupling matter Feynman diagrams in spin-foam models.

The Dirac determinant in Eq.(\ref{correlator3}) has been studied in the previous section. So we are going to find an expression for the inverse of the spin-foam Dirac operator $\sD^{-1}_{v_1,v_{2}}[j_f,i_e,g_{ve}]$. Here we put in a small regulator $\eps>0$ and consider the inverse matrix\footnote{In case $\sD^{-1}$ exists, $\lim_{\eps\to0}(\sD+\eps)^{-1}=\sD^{-1}$. Even in the case that $\sD^{-1}$ doesn't exist, recall the Feynman propagator
\be
\frac{i}{i\,\slash\!\!\!\partial-m+i\eps}=\frac{i}{\slash\!\!\!p-m+i\eps}
\ee
We see that the regulator $\eps$ corresponds to a Feynman regulator for the free quantum field.}
$(\sD+\eps)^{-1}$. Since the spin-foam Dirac operator $\sD$ is anti-Hermitian, the spectrum of operator $\sD+\eps$ lies in the right-half complex plane, thus $e^{-t(\sD+\eps)}$ $t\in[0,\infty)$ gives a contraction semigroup on a finite-dimensional vector space (with the usual norm). Then we have the following strong operator-equation as a consequence of the Hille-Yosida theorem \cite{reedsimon}:
\be
(\sD+\eps)^{-1}=\int_0^{\infty}\rmd L\ e^{-(\sD+\eps)L}
\ee
which is also known as Schwinger's proper time representation in physics literatures. Given two vertices $v,v'$
\be
(\sD+\eps)^{-1}_{v,v'}&=&\int_0^{\infty}\rmd L\ e^{-\eps L}\lt[e^{-\sD L}\rt]_{v,v'}\nonumber\\
&=&\int_0^{\infty}\rmd L\ e^{-\eps L}\sum_{\text{Path}_{v\to v'}}\frac{(-1)^k}{k!}{L^k}\sD_{v,v_1}\sD_{v_1,v_2}\cdots \sD_{v_{k-1},v'}\nonumber\\
&=&\int_0^{\infty}\rmd L\ e^{-\eps L}\sum_{\text{Path}_{v\to v'}}\frac{(-1)^k}{k!}{L^k}\prod_{i=1}^k(2V_{e_i})
\Big(g_{e_1 v}^\dagger g_{e_1v_1}\Big)
\Big(g_{e_2 v_1}^\dagger g_{e_2v_2}\Big)\cdots \Big(g_{e_k v_{k-1}}^\dagger g_{e_k v'}\Big)(-1)^{\text{Op}}\label{walk0}
\ee
where $\sD$ is nonvanishing only when $v,v'$ are neighboring vertices, $\sum_{\text{Path}_{v\to v'}}$ denotes the sum over all the oriented paths from $v$ to $v'$ along the edges of the 2-complex $\ck$, and the number $\text{Op}$ denote the number of edges on the path, whose orientation are opposite to the path.

For a Dirac fermion the regulator $\eps$ should be replaced by $iM_{v,v'}+\eps=i\delta_{v,v'}\fvol_v m_0+\eps$ where $\fvol_v$ is the volume of a 4-simplex at $v$ and $m_0$ is the fermion (bare) mass, in this case the inverse Dirac operator
\be
(\sD+iM+\eps)^{-1}_{v,v'}
&=&\int_0^{\infty}\rmd L\ e^{-(iM+\eps) L}\times\nonumber\\
&&\sum_{\text{Path}_{v\to v'}}\frac{(-1)^k}{k!}{L^k}\prod_{i=1}^k(2V_{e_i})
\Big(G_{v e_1 }\g^0 G_{e_1v_1}\Big)
\Big(G_{v_1 e_2 }\g^0 G_{e_2v_2}\Big)\cdots \Big(G_{v_{k-1}e_k}\g^0 G_{e_k v'}\Big)(-1)^{\text{Op}}\label{walk}
\ee

\subsection{World-line representation}

There is another representation of $(\sD+\eps)^{-1}_{v,v'}$ in terms of an discretized world-line action\footnote{On Minkowski spacetime, the discussion of the world-line representation of bosonic propagator often can be found in string theory textbooks, e.g. \cite{kiritsis}. For the fermionic propagator on Minkowski spacetime, the discussion of world-line representation can be found in e.g. \cite{worldline} and the reference therein.}, which is physically interesting. This representation is obtained by discretizing the exponential $e^{-\sD L}$
\be
(\sD+\eps)^{-1}_{v,v'}&=&\int_0^{\infty}\rmd L\ e^{-\eps L}\lt[e^{-\sD \frac{L}{n}\cdot n}\rt]_{v,v'}\nonumber\\
&=&\int_0^{\infty}\rmd L\ e^{-\eps L}\sum_{\text{Path}^n_{v\to v'}}\Big[e^{-\frac{L}{n}\sD}\Big]_{v,v_1}\Big[e^{-\frac{L}{n}\sD}\Big]_{v_1,v_2}\cdots \Big[e^{-\frac{L}{n}\sD}\Big]_{v_{n-1},v'}
\ee
In case the number of vertices of $\ck$ goes to be large
\be
&=&\int_0^{\infty}\rmd L\ e^{-\eps L}\lim_{n\to\infty}\sum_{\text{Path}^n_{v\to v'}}\Big[1-\frac{L}{n}\sD\Big]_{v,v_1}\Big[1-\frac{L}{n}\sD\Big]_{v_1,v_2}\cdots \Big[1-\frac{L}{n}\sD\Big]_{v_{n-1},v'}\nonumber\\
&=&\int_0^{\infty}\rmd L\ e^{-\eps L}\lim_{n\to\infty}\sum_{\text{Path}^n_{v\to v'}}
\Big[\delta_{v,v_1}-2\zeta^{e_1}_{v,v_1}V_{e_1}\frac{L}{n}g_{e_1 v}^\dagger g_{e_1v_1}\Big]
\Big[\delta_{v_1,v_2}-2\zeta^{e_2}_{v_1,v_2}V_{e_2}\frac{L}{n}g_{e_2 v_1}^\dagger g_{e_2v_2}\Big]\cdots\nonumber\\
&&\cdots
\Big[\delta_{v_{n-1},v'}-2\zeta^{e_n}_{v_{n-1},v'}V_{e_n}\frac{L}{n}g_{e_n v_{n-1}}^\dagger g_{e_n v'}\Big]
\ee
where $\text{Path}^n_{v\to v'}$ denotes the set of paths passing through $n-1$ vertices except $v$ and $v'$, $\zeta^e_{v,v'}=1$ if the orientation of $e$ coincide with $\overrightarrow{(v,v')}$ and $\zeta^e_{v,v'}=-1$ otherwise. We make a change of variable $L\mapsto\frac{L}{\ell_p^4}$ to make $L$ has a dimension of length/time
\be
&=&\frac{1}{\ell_p^4}\int_0^{\infty}\rmd L\ e^{-\frac{\eps}{\ell_p^4}L}\lim_{n\to\infty}\sum_{\text{Path}^n_{v\to v'}}
\Big[\delta_{v,v_1}-\zeta^{e_1}_{v,v_1}\frac{2 V_{e_1}}{\ell_p^3}\frac{l_1}{\ell_p}g_{e_1 v}^\dagger g_{e_1v_1}\Big]
\Big[\delta_{v_1,v_2}-\zeta^{e_2}_{v_1,v_2}\frac{2 V_{e_2}}{\ell_p^3}\frac{l_2}{\ell_p}g_{e_2 v_1}^\dagger g_{e_2v_2}\Big]\cdots\nonumber\\
&&\cdots
\Big[\delta_{v_{n-1},v'}-\zeta^{e_n}_{v_{n-1},v'}\frac{2 V_{e_n}}{\ell_p^3}\frac{l_n}{\ell_p}g_{e_n v_{n-1}}^\dagger g_{e_n v'}\Big]
\ee
where $l_1+l_2+\cdots+l_n=L$. Recall that
\be
g_{ev}^\dagger g_{ev'}=n_I(e)\sig^I=n_\a(e)\sig^\a(e)
\ee
where $n^\a(e)$ is the normalized tangent vector along the edge $e$ at the begin point $b(e)$, then $\zeta^e_{v,v'}n^\a(e)\equiv \mathbf{n}^\a(e)$ is the normalized tangent vector along the edge $\overrightarrow{(v,v')}$ at $b(e)$. Therefore in the limit $n\to\infty$
\be
(\sD+\eps)^{-1}_{v,v'}&=&\frac{1}{\ell_p^4}\int_0^{\infty}\rmd L\ e^{-\frac{\eps}{\ell_p^4}L}\sum_{\g\in\text{Path}_{v\to v'}}\cp\exp\lt[-\frac{2}{\ell_p^4}\int_0^L\rmd l\ \mathbf{n}_\a\Big(\g(l)\Big)\sig^\a\Big(\g(l)\Big)V\Big(\g(l)\Big)\rt]
\ee
where $\mathbf{n}_\a(\g(l))$ is the normalized tangent vector along the path $\g(l)$, $V(\g(l))$ is the 3-volume of the tetrahedron $\t(l)$ at $\g(l)$. We then make a change of variable and define $t=l/L$
\be
(\sD+\eps)^{-1}_{v,v'}&=&\frac{1}{\ell_p^4}\int_0^{\infty}\rmd L\ e^{-\frac{\eps}{\ell_p^4}L}\sum_{\g\in\text{Path}_{v\to v'}}\cp\exp\lt[-\frac{2}{\ell_p^4}\int_0^1\rmd t\ L\ \mathbf{n}_\a\Big(\g(t)\Big)\sig^\a\Big(\g(t)\Big)V\Big(\g(t)\Big)\rt]\
\ee
This expression can be written as a gauge-fixed path integral ($\mathbf{n}^\a(t)=\mathbf{n}^\a(\g(t))$, $\sig^\a(t)=\sig^\a(\g(t))$)
\be
&=&\frac{1}{\ell_p^4}\int_0^{\infty}\rmd L\int [De(t)]\int_{\g(0)=v}^{\g(1)=v'} [D\g(t)]\prod_{t\in[0,1]} \delta(e(t)-L)\ e^{-\frac{\eps}{\ell_p^4}\int_0^1\rmd t\ e(t)} \cp e^{-\frac{2}{\ell_p^4}\int_0^1\rmd t\ e(t) \mathbf{n}_\a(t)\sig^\a(t)V(t)}
\ee
where $\int_{\g(0)=v}^{\g(1)=v'} [D\g(t)]=\sum_{\text{Path}_{v\to v'}}$ is nothing but a counting measure. If we define a world-line action (matrix)\footnote{Actually it can also be considered as a ``world-tube''action because of the 3-volume $V(t)$ i.e.
\be
\mathbf{S}_W=\frac{2}{\ell_p^4}\int_0^1\rmd t\ e(t)\int_{\t(t)}\rmd^3x\sqrt{\det q}\lt[ \mathbf{n}_\a(t)\sig^\a(t)+\frac{1}{2}\eps'\rt]
\ee
where we redefine the regulator $\eps'$ such that $\eps=\eps'V(t)$.}
\be
\mathbf{S}_W[e,\g,j_f,i_e,g_{ve}]=\frac{2}{\ell_p^4}\int_0^1\rmd t\lt[ e(t) \mathbf{n}_\a(t)\sig^\a(t)V(t)+\frac{1}{2}\eps\ e(t)\rt]
\ee
and consider $e(t)$ as a world-line metric. Then the inverse spin-foam Dirac operator is a discretized path integral of this world-line action on the spin-foam background
\be
(\sD+\eps)^{-1}_{v,v'}=\frac{1}{\ell_p^4}\int_0^{\infty}\rmd L\int [De(t)]\int_{\g(0)=v}^{\g(1)=v'} [D\g(t)]\prod_{t\in[0,1]} \delta(e(t)-L)\ \cp e^{-\mathbf{S}_W[e,\g,j_f,i_e,g_{ve}]}
\ee
Obviously, $\prod_{t\in[0,1]} \delta(e(t)-L)$ is a gauge-fixing for the world-line reparametrization invariance, and the parameter $L$ is a world-line Teichm\"uller parameter, which is a gauge fixing left-over.

For Dirac fermion, the fermion world-line action reads
\be
\mathbf{S}_D[e,\g,j_f,i_e,g_{ve}]=\frac{2}{\ell_p^4}\int_0^1\rmd t\lt[ e(t) \mathbf{n}_\a(t)\g^\a(t)V(t)+\frac{1}{2}\lt(im_0\fvol(t)+\eps\rt) e(t)\rt]
\ee
where the discretized version of $\mathbf{n}_\a(t)\g^\a(t)$ is $\zeta^e_{v,v'}G_{ve}\g^0G_{ev'}$. Then the inverse spin-foam Dirac operator is a discretized version of world-line path integral
\be
(\sD+iM+\eps)^{-1}_{v,v'}=\frac{1}{\ell_p^4}\int_0^{\infty}\rmd L\int [De(t)]\int_{\g(0)=v}^{\g(1)=v'} [D\g(t)]\prod_{t\in[0,1]} \delta(e(t)-L)\ \cp e^{-\mathbf{S}_D[e,\g,j_f,i_e,g_{ve}]}.
\ee

\section{Conclusion and Discussion}

We have defined and discussed the fermion quantum field coupled with spin-foam quantum gravity, we have defined and explored the properties of the fermion correlation functions on spin-foams, where we have shown there is a spin-foam analog of PCT symmetry for spin-foam fermions. The concrete evaluation of the fermion correlations function has also been performed, and the main building blocks, the Dirac determinant and the inverse Dirac operator, has been computed. We have shown that the spin-foam fermion correlation functions can be represented as the Feynman diagrams of fermion world-lines imbedded in the spin-foam amplitudes. In this article we have considered only the interaction between fermions and gravity, so the Feynman diagram imbedded in the spin-foams are factorized into disconnected propagators. We expect that even for interacting matter quantum fields, a similar structure holds, i.e. the matter field correlation functions could be represented (at least perturbatively) by Feynman diagrams of the interacting fields imbedded in the spin-foam amplitudes.

In closing, we present a remark about the species doubling problem for lattice fermions. The spin-foam fermions are defined with a discrete setting, similar to the fermions in lattice field theory. It is well-known that the formal discretization of the fermion action on a lattice suffers the problem of species doubling (see any textbook on lattice field theory e.g. \cite{lattice}, see also \cite{anomaly}), while the problem is resolved when the discretized Dirac operator satisfies the Ginsparg-Wilson relation. Such a Dirac operator can be constructed from the formal discretization in an overlap formulation (overlap fermions) by Neuberger \cite{overlap}, which gives an exact chiral symmetry and anomaly calculation (see \cite{anomaly} for a summary). The overlap fermions can also be defined on a curved lattice in the presence of external gravitational field \cite{overlapG}, where the chiral symmetry and anomaly calculation are reproduced correctly. Because of these results, one might consider if the overlap formulation should be employed to define the spin-foam fermion, instead of the formal discretization used in the present work. First of all such an idea could be realized straight-forwardly in the formulation of overlap fermion, following the technique for example in \cite{overlapG}. However the overlap formulation (and Ginsparg-Wilson relation) make correction for the formal discretized Dirac operator by additional terms proportional (and higher order of) the lattice spacing $a$, which is a semiclassical concept in the context of spin-foam model. Thus it seems to us that it is unnatural to implement those corrections fundamentally. But it would be interesting to see if those corrections can emerge from some certain semiclassical approximations of spin-foam model. It is not hopeless in our opinion for the following reasons: the summing over all the geometries in the spin-foam model make it hopeful that the fermion doublers are canceled in a similar way to those on a flat random lattice. Some evidences for this have been shown in the context of fermion on Regge gravity \cite{Ren}, where the fermion propagator are computed numerically and display excellent agreement with the continuum field theory.

\section*{Acknowledgments}

The authors would like to thank E. Bianchi, E. Magliaro, E. Livine, C. Perini, and W. Wieland for fruitful discussions. M.H. would also like to thank Song He for discussions and his comments from a lattice-field-theorist's perspective.

\end{document}